\documentclass[conference,compsoc]{IEEEtran}
\usepackage{balance}
\usepackage{flushend}
\usepackage{graphicx}
\usepackage{subfigure}
\usepackage{algorithm}
\usepackage{algpseudocode}
\usepackage{amsmath}
\usepackage{bbm}
\usepackage{amsthm}
\usepackage{amsfonts}
\usepackage{stfloats}
\usepackage{url}
\usepackage{xspace}
\usepackage{multirow}
\usepackage{booktabs}
\usepackage{mathtools}
\usepackage{caption}
\usepackage{hyperref} 
\usepackage{color}

\usepackage{multirow}
\usepackage{bbding}
\usepackage{colortbl}
\usepackage{textcomp}
\usepackage{comment}
\usepackage{enumerate}
\usepackage{tcolorbox}
\usepackage{pifont}
\usepackage{cite}
\usepackage[numbers,sort&compress]{natbib}

\usepackage{enumitem}
\usepackage{tabularx}
\usepackage{hyperref}

\usepackage{array}
\usepackage{ragged2e}
\usepackage{longtable}

\setlist[itemize]{leftmargin=*}

\usepackage{xcolor}
\hypersetup{
  colorlinks,
  linkcolor={blue!70!green},
  citecolor={green!70!blue},
  urlcolor={orange!70!red}
}

\usepackage{hyperref}
\makeatletter
\def\UrlAlphabet{%
      \do\a\do\b\do\c\do\d\do\e\do\f\do\g\do\h\do\i\do\j%
      \do\k\do\l\do\m\do\n\do\o\do\p\do\q\do\r\do\s\do\t%
      \do\u\do\v\do\w\do\x\do\y\do\z\do\A\do\B\do\C\do\D%
      \do\E\do\F\do\G\do\H\do\I\do\J\do\K\do\L\do\M\do\N%
      \do\O\do\P\do\Q\do\R\do\S\do\T\do\U\do\V\do\W\do\X%
      \do\Y\do\Z}
\def\UrlDigits{\do\1\do\2\do\3\do\4\do\5\do\6\do\7\do\8\do\9\do\0}
\g@addto@macro{\UrlBreaks}{\UrlOrds}
\g@addto@macro{\UrlBreaks}{\UrlAlphabet}
\g@addto@macro{\UrlBreaks}{\UrlDigits}
\makeatother

\definecolor{mygray}{gray}{.9}

\newcommand{\etal}{\textit{et al.}\xspace}
\newcommand{\ie}{\textit{i.e.}\xspace}
\newcommand{\eg}{\textit{e.g.}\xspace}

\newcommand{\mypara}[1]{\smallskip\noindent\textbf{#1.} \xspace}

\definecolor{revision}{RGB}{0,0,0}
\newcommand{\rev}[1]{{\color{revision} #1\xspace}\xspace}
\newcommand{\revstart}{\begin{color}{revision}}
\newcommand{\revend}{~\!\!\end{color}}

\definecolor{new content}{RGB}{0,255,0}

\usepackage{etoolbox}
\makeatletter
\patchcmd{\hyper@makecurrent}{%
    \ifx\Hy@param\Hy@chapterstring
        \let\Hy@param\Hy@chapapp
    \fi
}{%
    \iftoggle{inappendix}{%
        \@checkappendixparam{chapter}%
        \@checkappendixparam{section}%
        \@checkappendixparam{subsection}%
        \@checkappendixparam{subsubsection}%
        \@checkappendixparam{paragraph}%
        \@checkappendixparam{subparagraph}%
    }{}%
}{}{\errmessage{failed to patch}}

\newcommand*{\@checkappendixparam}[1]{%
    \def\@checkappendixparamtmp{#1}%
    \ifx\Hy@param\@checkappendixparamtmp
        \let\Hy@param\Hy@appendixstring
    \fi
}
\makeatletter

\newtoggle{inappendix}
\togglefalse{inappendix}

\apptocmd{\appendix}{\toggletrue{inappendix}}{}{\errmessage{failed to patch}}

\usepackage{tikz}
\usepackage{oplotsymbl}

\DeclareRobustCommand{\IEEEauthorrefmark}[1]{\smash{\textsuperscript{\footnotesize #1}}}
\IEEEoverridecommandlockouts

\tcbuselibrary{breakable} %

\begin{document}

\title{SoK: Dataset Copyright Auditing in Machine Learning Systems}
\author{\IEEEauthorblockN{Linkang Du\IEEEauthorrefmark{1}\IEEEauthorrefmark{*} %
Xuanru Zhou\IEEEauthorrefmark{2}\IEEEauthorrefmark{*} \qquad
Min Chen\IEEEauthorrefmark{3} \qquad
Chusong Zhang\IEEEauthorrefmark{2} \\
Zhou Su\IEEEauthorrefmark{1} \qquad
Peng Cheng\IEEEauthorrefmark{2} \qquad
Jiming Chen\IEEEauthorrefmark{2}$^,$\IEEEauthorrefmark{4} \qquad
Zhikun Zhang\IEEEauthorrefmark{2}\IEEEauthorrefmark{\#}
}
\IEEEauthorblockA{\IEEEauthorrefmark{1}Xi'an Jiaotong University \quad \IEEEauthorrefmark{2}Zhejiang University \quad \IEEEauthorrefmark{3}Vrije Universiteit Amsterdam  \quad \IEEEauthorrefmark{4}Hangzhou Dianzi University}
}

\maketitle

\def\thefootnote{*}\footnotetext{The first two authors made equal contribution.}
\def\thefootnote{\#}\footnotetext{Zhikun Zhang is the corresponding author.}
\renewcommand*{\thefootnote}{\arabic{footnote}}

\begin{abstract}
As the implementation of machine learning (ML) systems becomes more widespread, especially with the introduction of larger ML models, we perceive a spring demand for massive data. 
However, it inevitably causes infringement and misuse problems with the data, such as using unauthorized online artworks or face images to train ML models. 
To address this problem, many efforts have been made to audit the copyright of the model training dataset. 
However, existing solutions vary in auditing assumptions and capabilities, making it difficult to compare their strengths and weaknesses.
In addition, robustness evaluations usually consider only part of the ML pipeline and hardly reflect the performance of algorithms in real-world ML applications. 
Thus, it is essential to take a practical deployment perspective on the current dataset copyright auditing tools, examining their effectiveness and limitations. 
Concretely, we categorize dataset copyright auditing research into two prominent strands: \textit{intrusive} methods and \textit{non-intrusive} methods, depending on whether they require modifications to the original dataset. 
Then, we break down the intrusive methods into different watermark injection options and examine the non-intrusive methods using various fingerprints. 
To summarize our results, we offer detailed reference tables, highlight key points, and pinpoint unresolved issues in the current literature. 
By combining the pipeline in ML systems and analyzing previous studies, we highlight several future directions to make auditing tools more suitable for real-world copyright protection requirements. 
\end{abstract}

\section{Introduction}
Deep neural network (DNN) models, as a promising ML paradigm, are becoming an increasingly ubiquitous part of our daily lives~\cite{Canziani2016AnAO, Wang2023ResilientDC, Liu2017ASO, Sze2017EfficientPO, Samek2021ExplainingDN, Miikkulainen2017EvolvingDN, Zhang2023VulnerabilityOM, Ren2024StealthyBA}. 
DNN models are eager for large-scale datasets, as they benefit from abundant training examples to learn complex patterns and representations. 
Notable examples, such as GPT-4~\cite{openai2023gpt4}, T5~\cite{raffel2020exploring}, CLIP~\cite{RKHRGASAMCKS21}, DALL·E 3~\cite{betker2023improving}, and AlphaZero~\cite{silver2017mastering}, demonstrate the power of training with large datasets. 
These models show the impact of large datasets in pushing the boundaries of AI capabilities. 

Large-scale datasets that pave the way for real-world ML systems also open the door to potential data misuse and abuse. 
In 2020, Kashmir Hill from The New York Times brought to light the potential risks associated with the misuse of facial data. 
She focused on Clearview.AI, a private company that amassed over 3 billion images from ``public sources'' to create a facial recognition system~\cite{Hill.2021}. 
This system is capable of identifying hundreds of millions of individuals without their knowledge or consent. 
Insider attacks, such as data theft by departing employees, are also a major cause of data infringement. 
Tessian reported that 40\% of US employees took their generated data with them when leaving their jobs~\cite{TessianReport}. 
Similarly, a 2021 survey~\cite{BiscomReport} found that more than a quarter of the respondents admitted to taking data upon departure, with 95\% attributing this theft to a lack of policies or technologies designed to prevent data theft by exiting employees. 

To prevent misuse of copyrighted datasets, the technique of \textit{dataset copyright auditing} has gained substantial attention from both the academic and industrial community~\cite{DBLP:conf/icml/SablayrollesDSJ20, DBLP:conf/iclr/MainiYP21, DBLP:journals/corr/abs-2303-11470, DBLP:journals/corr/abs-2209-06015, DU2024WIP, zhu2023detection}. 
The goal is to determine whether a suspicious model was created by unauthorized entities who illegally gained access to the copyrighted dataset. 
The state-of-the-art dataset copyright auditing strategies vary significantly in their assumptions and techniques, each adapts to different application scenarios. 
Therefore, it is essential to holistically understand their similarities, highlight their performance trade-offs, and enlighten future research paths.

\mypara{Our Contributions}
In this paper, we systematize the state-of-the-art research on dataset copyright auditing and categorize existing work according to different technical principles. 
We focus on the potential issues of existing approaches that have been deployed in practice. 
Considering the influential factors in practice, we benchmark existing technique routes and summarize a series of observations, open problems, and future directions. 
Concretely, we make the following contributions. 
\begin{itemize}    
    \item \textbf{We compare existing solutions based on their application scope, technique used, required authority, and evaluation settings. }
    The existing solutions for dataset copyright auditing encompass the full range of dataset granularities, from an individual sample to the entire dataset. 
    Most current methods are specifically designed for auditing image data in classification tasks. 
    Membership inference~\cite{SSSS17, LZ21, liu2022membership} and backdoor~\cite{GDG17, CLLLS17} techniques are the most commonly employed foundational components of these solutions. 
    Regarding the required level of access, the majority of existing work can perform copyright auditing through black-box interaction with the suspect model without the need for an auxiliary dataset. 
    In addition to auditing effectiveness, aspects such as stealthiness and robustness are also evaluated in existing studies.
    
    \item \textbf{We divide existing methods into two categories and five subcategories, and summarize five key takeaways along with six open problems for practitioners. }
    Based on whether modifications to the original dataset are required, we first classify existing work into two paradigms: intrusive auditing and non-intrusive auditing. 
    The first category involves altering the original dataset, such as embedding backdoors or adding spurious features. 
    The second category includes approaches that do not require any modifications to the dataset. 
    We then analyze these two paradigms separately to highlight the strengths and limitations of current technical approaches. 
    For example, within backdoor-based auditing methods, those based on targeted backdoors often exhibit better auditing effectiveness, yet their stealthiness may not be as good as methods based on untargeted backdoors. 
    
    \item \textbf{We delve into the pipeline of ML systems and explore potential factors that may affect the auditing effectiveness in the wild. }
    We partition the process of developing the ML system into three stages: data preparation, model training, and model deployment. 
    By analyzing robustness evaluations in the existing literature and commonly used processing strategies in practice, we construct two practical scenarios and three adversarial scenarios to evaluate the effectiveness, stealthiness, and side effects (impact on the normal performance of the model) of different technical approaches. 
    For example, we observe that targeted backdoor-based auditing methods demonstrate strong robustness in all evaluated cases and maintain auditing performance even when DP-SGD~\cite{ACGMMTZ16} severely degrades the model performance. 
    
    \item \textbf{We identify several unresolved problems and challenges for future research. }
    First, factors in \textit{data preparation}, \textit{model training}, and \textit{model deployment} within the ML pipeline affect the effectiveness of auditing algorithms. 
    Although numerous studies explore robustness across these stages, a comprehensive assessment integrating all three remains rare, underscoring the need for a holistic evaluation toolbox for dataset copyright auditing. 
    Second, existing auditing solutions predominantly focus on image classification. 
    Although some advancements have been made in copyright auditing schemes for other domains, such as text~\cite{Liu2023WatermarkingCD, SSWM20, li2023functionmarker}, audio~\cite{DBLP:journals/corr/abs-2303-11470, Chen2023WavMarkWF, Natgunanathan2017PatchworkBasedMA}, or large language models (LLMs)~\cite{yao2023promptcare, li2024digger}, insufficient attention has been paid to the dataset copyright issues posed by LLMs and multi-modal models.
    In addition, current auditing approaches primarily evaluate algorithm performance based on accuracy metrics. 
    Future work should consider providing formal guarantees in copyright verification, which is critical in legal contexts. 
\end{itemize}

\mypara{Roadmap}
We first formalize the dataset copyright auditing problem and provide an overview of the existing solutions in~\autoref{sec:dataset-auditing-motivation-and-threat-Model}. 
Next, we conduct a detailed analysis of existing dataset copyright auditing strategies, \ie, the intrusive auditing paradigm (in~\autoref{sec:intrusive-auditing}) and the non-intrusive auditing paradigm (in~\autoref{sec:non-intrusive-auditing}). 
In these sections, we describe the core operations in dataset copyright auditing, existing works, and a summary of the takeaways and open problems for each subcategory. 
Then, involved with the ML system pipeline, we perform an analysis of dataset copyright auditing strategies from a practical perspective in~\autoref{sec:copyright-auditing-in-the-wild}, and discuss the promising directions in the future development of dataset copyright auditing in~\autoref{sec:promising-directions-in-future}. 

\section{Dataset Copyright Auditing}
\label{sec:dataset-auditing-motivation-and-threat-Model}

\subsection{Problem Statement}
\label{sec:problem-statement}
\begin{figure}[!t]
\centering
\includegraphics[width=\hsize]{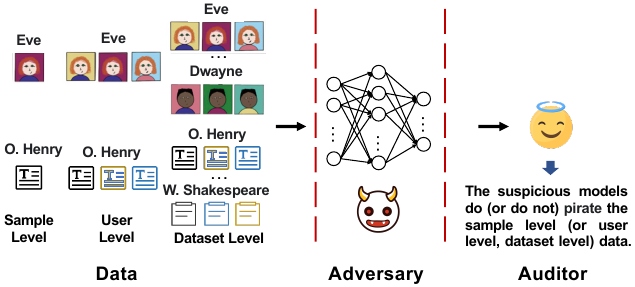}
\caption{A typical application scenario of the existing dataset copyright auditing mechanisms. 
}
\label{fig:application_scenario}
\end{figure}

\mypara{Application Scenarios}
We formalize the classic auditing scenario, where two participants exist, \ie, an owner of the dataset (owner) and a suspicious model (adversary). 
The owner collects and then publishes the dataset online or sells the dataset to others. 
The adversary with access to the dataset illegally trains and makes profits from ML models. 
\autoref{fig:application_scenario} illustrates an application example in which an owner shares its data with a third party, such as posting personal photos on Instagram or expressing opinions on Twitter~\footnote{The hand-drawn images used in this paper are credited to \href{https://www.emilywenger.com/}{Emily Wenger} at Duke University, under the \href{https://creativecommons.org/licenses/by/4.0/}{CC4.0} license.}. 
However, a malicious company (adversary) with access to the data illegally builds a Model-as-a-Service (MaaS) platform, and then profits from the MaaS platform or infringes on the user's portrait rights~\cite{NYTIMESH20, NYTIMESH21}. 
The owner suspects that the ML models are generated by its data and thus can leverage the auditing tools to determine whether the adversary pirates their private data. 
Existing strategies are usually designed for a specific granularity of the data, \ie, the sample level, the user level, and the dataset level. 
``sample level'' refers to individual data points, ``user level'' pertains to aggregated data from individual users or subjects, 
and ``dataset level'' encompasses the entire collection of data, assessing overall characteristics and performance of the model across all samples and users. 
Each level offers a different perspective, from the granular details of single instances to the broad overview of the entire dataset. 
For example, FACE-AUDITOR~\cite{CZWBZ23} is designed to audit datasets at the user level, while RAI2~\cite{DBLP:conf/ndss/DongLCXZ023} is targeted for auditing at the dataset level. 
The fine-grained auditing methods can be extended to coarse-grained auditing scenarios. For example, if the dataset owner discovers that a model utilizes more than a certain predefined proportion (\eg, 80\%) of the dataset samples, a claim of dataset-level infringement can be made. 

\mypara{Auditor's Background Knowledge and Capability}
Existing work usually assumes that the owner of the dataset has full access to its dataset \ie, \textit{the target dataset}. 
Regarding access to the suspicious model, some studies~\cite{DBLP:journals/corr/abs-2303-11470, DBLP:conf/nips/LiBJYXL22, DBLP:conf/icml/Choquette-ChooT21} only use a set of inputs to obtain the corresponding outputs of the suspicious model. 
This is known as ``black-box access'' and is the most general and challenging audit scenario. 
Some studies~\cite{DBLP:conf/icml/SablayrollesDSJ20, DBLP:journals/corr/abs-2209-06015, Xu2022DataOI} examine the impact of having white-box access to the suspicious model for audit purposes, such as knowing the structure and parameters of the model. 

\mypara{Dataset Copyright Auditing Problem}
The dataset copyright auditing problem $\mathcal{A}$ can be defined as follows: 
\begin{equation}
\mathcal{A}: g(x, f(x)) \rightarrow \text{0 or 1}, 
\label{eq:dataset-auditing-problem-simple}
\end{equation}
where $x$ is the target dataset, and $f$ is the suspicious model. 
The dataset owner utilizes the auditing method $g$ to check whether the suspicious model infringes its dataset. 
If dataset infringement occurs, $g$ predicts 1; otherwise, it outputs 0. 

\begin{table*}[h]
\caption{An overview of existing dataset copyright auditing methods. 
\textbf{Data}: Sample, User, and Dataset in the Protection Level column indicate for what auditing granularity the methods are optimized. 
D1-D5 in the Domain column represent the image, text, tabular, audio, and graph data, respectively. 
C, R, and RL in the Task column represent the abbreviations of the \underline{c}lassification, \underline{r}egression, and \underline{r}einforcement \underline{l}earning tasks, respectively. 
\textbf{Tech.}: MI, DI, B, CT, and SF represent the abbreviations of membership inference, dataset inference, backdoor, color transformation, and spurious features. 
\textbf{Access}: \circletfill=black-box, \circlet=white-box, \circletfillhl=both white-box and black-box. 
\textbf{Dataset Modification}: \circletfill=sample x and label y, \circletfillhl=sample x or label y, \circlet=Neither. 
\textbf{Auxiliary Dataset}: \ding{51}=essential, \ding{55}=non-essential. 
\textbf{Stealthiness Testing}: \ding{51}=Assessed the stealthiness of the methods that require modification of the original dataset. \ding{55}=Not assessed the stealthiness of methods that require the modification of the original dataset. 
/=inapplicable. 
\textbf{Robustness Testing}: S1, S2, and S3 represent the abbreviations of the main steps in the machine learning pipeline, \ie, the data preparation process, the model training process, and the model deployment phase. 
\textbf{Code}: whether the paper open-sources its code (\ding{51}) or not (\ding{55}). 
}
\label{tab:overview-of-existing-work}
\centering
\small
\setlength{\tabcolsep}{0.6em}
\renewcommand{\arraystretch}{1.1}
\begin{tabular}{ccccccccccc} 
\toprule
\multicolumn{3}{c}{\textbf{Data}}                      & \textbf{Tech.} & \multicolumn{3}{c}{\textbf{Auditing Cost}}                                                                                                         & \multicolumn{2}{c}{\textbf{Evaluation}}                                                                                                                                                          & \textbf{Code} & \textbf{Reference}                       \\ 
\cline{1-3}\cline{5-9}
\begin{tabular}[c]{@{}c@{}}Protection\\Level\end{tabular}                              & Domain     & Task &                & Access                & \begin{tabular}[c]{@{}c@{}}Dataset\\Modification\end{tabular} & \begin{tabular}[c]{@{}c@{}}Auxiliary\\Dataset\end{tabular}  & Stealthiness & Robustness &               &                                           \\ 
\toprule
\multirow{9}{*}{\textbf{Sample}}   & D1         & C    & MI             & \circletfill\xspace   & \circlet\xspace                                               & \ding{51}                                                   & /                                                             & S2, S3                                                      & \ding{51}     & \cite{SSSS17}                             \\
                                   & D1         & C    & MI             & \circletfill\xspace   & \circlet\xspace                                               & \ding{51}                                                   & /                                                             & S1                                                          & \ding{55}     & \cite{SDSOJ19}                            \\
                                   & D1         & C    & MI             & \circletfill\xspace   & \circlet\xspace                                               & \ding{55}                                                   & /                                                             & S2                                                          & \ding{55}     & \cite{LF20}                               \\
                                   & D1         & C    & DI             & \circletfill\xspace   & \circlet\xspace                                               & \ding{55}                                                   & /                                                             & S2, S3                                                      & \ding{51}     & \cite{DBLP:conf/icml/Choquette-ChooT21}   \\
                                   & D1         & C    & B              & \circletfill\xspace   & \circlet\xspace                                               & \ding{51}                                                   & /                                                             & S2, S3                                                      & \ding{51}     & \cite{LZ21}                               \\
                                   & D1         & C    & CT             & \circletfill\xspace   & \circletfillhl\xspace                                         & \ding{55}                                                   & \ding{51}                                                     & S1                                                          & \ding{51}     & \cite{DBLP:conf/eccv/ZouGW22}             \\
                                   & D1, D3     & C    & MI             & \circletfill\xspace   & \circlet\xspace                                               & \ding{55}                                                   & /                                                             & S3                                                          & \ding{51}     & \cite{liu2022membership}                  \\ 
                                   & D2         & C    & MI \& B          & \circletfill\xspace   & \circletfillhl\xspace                                         & \ding{55}                                                   & \ding{51}                                                     & S3                                                          & \ding{51}     & \cite{Liu2023WatermarkingCD}              \\
                                   & D3         & RL   & MI             & \circletfill\xspace   & \circlet\xspace                                               & \ding{55}                                                   & /                                                             & S2, S3                                                      & \ding{51}     & \cite{DCSJCCZ24, DCSJCCZ24arXiv}                          \\ 
\midrule
\multirow{4}{*}{\textbf{User}}     & D1         & C    & MI \& B          & \circletfill\xspace   & \circletfill\xspace                                           & \ding{55}                                                   & /                                                             & S2, S3                                                      & \ding{51}     & \cite{DBLP:conf/ijcai/HuSDCSZ22}          \\
                                   & D1         & C    & SF             & \circletfill\xspace   & \circletfillhl\xspace                                         & \ding{51}                                                   & \ding{51}                                                     & S2, S3                                                      & \ding{51}     & \cite{DBLP:journals/corr/abs-2208-13893}  \\
                                   & D1         & C    & MI             & \circletfill\xspace   & \circlet\xspace                                               & \ding{51}                                                   & /                                                             & S1, S2, S3                                                  & \ding{51}     & \cite{CZWBZ23}                            \\
                                   & D1, D2     & C    & MI             & \circletfill\xspace   & \circletfill\xspace                                           & \ding{55}                                                   & /                                                             & S3                                                          & \ding{51}     & \cite{SSWM20}                             \\ 
\midrule
\multirow{10}{*}{\textbf{Dataset}} & D1         & C    & B              & \circletfill\xspace   & \circletfill\xspace                                           & \ding{55}                                                   & \ding{55}                                                     & None                                                        & \ding{51}     & \cite{DBLP:journals/corr/abs-2010-05821}  \\
                                   & D1         & C    & SF             & \circletfillhl\xspace & \circletfillhl\xspace                                         & \ding{55}                                                   & \ding{51}                                                     & S1, S3                                                      & \ding{55}     & \cite{DBLP:conf/icml/SablayrollesDSJ20, Tekgul2022OnTE, tekgul2022effectiveness}   \\
                                   & D1         & C    & B              & \circletfill\xspace   & \circletfill\xspace                                           & \ding{55}                                                   & \ding{51}                                                     & S3                                                          & \ding{51}     & \cite{DBLP:conf/nips/LiBJYXL22}           \\
                                   & D1         & C    & DI             & \circletfillhl\xspace & \circlet\xspace                                               & \ding{55}                                                   & /                                                             & None                                                        & \ding{51}     & \cite{DBLP:conf/iclr/MainiYP21}           \\
                                   & D1         & C    & DI             & \circletfillhl\xspace & \circlet\xspace                                               & \ding{51}                                                   & /                                                             & S3                                                          & \ding{55}     & \cite{DBLP:journals/corr/abs-2209-09024}  \\
                                   & D1         & C    & SF             & \circletfill\xspace   & \circletfillhl\xspace                                         & \ding{55}                                                   & \ding{51}                                                     & S3                                                          & \ding{51}     & \cite{guo2023domain}                      \\
                                   & D2         & R    & SF             & \circletfill\xspace   & \circletfillhl\xspace                                         & \ding{55}                                                   & \ding{51}                                                     & S1                                                          & \ding{51}     & \cite{li2023functionmarker}                      \\
                                   & D1, D2     & C    & DI             & \circletfill\xspace   & \circlet\xspace                                               & \ding{51}                                                   & /                                                             & S3                                                          & \ding{51}     & \cite{DBLP:conf/ndss/DongLCXZ023}         \\
                                   & D1, D2, D3 & C, R & DI             & \circlet\xspace       & \circlet\xspace                                               & \ding{51}                                                   & /                                                             & None                                                        & \ding{51}     & \cite{Xu2022DataOI}                       \\
                                   & D1, D2, D4 & C    & B              & \circletfill\xspace   & \circletfillhl\xspace                                         & \ding{55}                                                   & \ding{51}                                                     & None                                                        & \ding{55}     & \cite{DBLP:journals/corr/abs-2303-11470}  \\
                                   & D1, D2, D5 & C    & B              & \circletfill\xspace   & \circletfill\xspace                                           & \ding{55}                                                   & \ding{51}                                                     & S3                                                          & \ding{51}     & \cite{DBLP:journals/corr/abs-2209-06015, li2023black}  \\
\bottomrule
\end{tabular}
\end{table*}

\subsection{Methodology}
\label{sec:methodology}
This paper primarily explores dataset copyright auditing strategies within deep learning contexts. 
We perform a literature search using the keywords ``data copyright'', ``data ownership'', ``data watermarking'', and ``data inference''. 
Given that membership inference can address copyright auditing, we include studies~\cite{SSSS17, LZ21, liu2022membership} in \autoref{tab:overview-of-existing-work}, which lists a total of 27 papers. 
Most of these articles were published after 2017, with a significant increase following the Clearview.AI controversy in 2020. 
We omit works focused on copyright protection for deep learning models~\cite{chen2022copy, krauss2024clearstamp}, instead we focus on the copyright of datasets. 
We empirically analyze existing methods from four aspects with a focus on the practical utility of these methods. 

We do not restrict our search to the image domain in the paper collection. 
However, \autoref{tab:overview-of-existing-work} shows that almost all the papers focus on the image domain, with a subsequent attention to text~\cite{Liu2023WatermarkingCD, SSWM20, li2023functionmarker} and audio~\cite{DBLP:journals/corr/abs-2303-11470}. 
Text domain strategies primarily utilize backdoor techniques for watermark design, whereas audio domain approaches often embed watermarks in the frequency domain to facilitate data traceability. 

\subsection{Overview of Existing Solutions}
\label{sec:overview-of-existing-works}
We provide an overview of existing dataset copyright auditing methods in \autoref{tab:overview-of-existing-work}. 
``Data'' refers to the scope of the application, 
``Tech.'' refers to the specific technique used, 
``Auditing Cost'' refers to the authority required during the audit process,
and ``Evaluation'' refers to whether the paper considers essential practical factors in the evaluation. 
From \autoref{tab:overview-of-existing-work}, we gain the following observations. 

\mypara{Observation 1: The existing solutions cover all the granularities of the dataset}
Zou~\etal~\cite{DBLP:conf/eccv/ZouGW22} proposed the first strategy optimized for the sample-level dataset copyright auditing. 
Before that, membership inference strategies against deep learning models can be adjusted to fit the sample-level dataset copyright auditing. 
Recently, dataset copyright auditing technologies at the user-level~\cite{SSWM20, CZWBZ23} and dataset-level~\cite{DBLP:conf/icml/SablayrollesDSJ20, DBLP:journals/corr/abs-2010-05821} have flourished with society’s increased attention on dataset copyright. 

\mypara{Observation 2: Existing solutions mainly aim at the classification task}
This observation is derived mainly from the ``Task'' column of \autoref{tab:overview-of-existing-work}. 
The reason is that current auditing strategies usually rely on the model's posterior probability or predicted labels~\cite{DBLP:conf/icml/Choquette-ChooT21, LZ21, DBLP:conf/iclr/MainiYP21, szyller2022robustness, tian2023knowledge, DBLP:conf/icml/Choquette-ChooT21} to determine whether the data were used in the model's training process. 
However, for the regression task, the model's posterior probability or predicted labels are difficult to define, which hinders the design of the auditing mechanisms.  

\mypara{Observation 3: More solutions designed for the image domain compared to other domains}
The image-based applications are popular in the real world, such as face recognition systems and art style transfer. 
In addition, the technologies that the auditing method relies on, such as backdoor or membership inference, have been widely studied in the field of image and can be easily integrated into the audit method.
For instance, Li~\etal\cite{DBLP:journals/corr/abs-2010-05821} adopted classical poisoning-based backdoor attacks, \eg BadNets~\cite{GDG17}, to watermark the image samples. 
Similarly, Wenger~\etal~\cite{DBLP:journals/corr/abs-2208-13893} chose to insert a spurious feature into images as watermark. 

\mypara{Observation 4: The majority can conduct the dataset copyright auditing with black-box access to the suspicious model}
According to the Access column in \autoref{tab:overview-of-existing-work}, most existing works assume that the dataset owner has only black-box access to the suspicious model. 
In this case, the auditor queries the suspicious model with a set of inputs to obtain the corresponding outputs. 
The auditor then analyzes the relationship between inputs and outputs to make a decision. 
Additionally, three works consider the scenario where the dataset owner has white-box access to the suspicious model. 
Maini~\etal~\cite{DBLP:conf/iclr/MainiYP21} proposed a dataset copyright auditing method based on the distances between the samples and the decision boundary of the suspicious model. 
They estimate the minimum distance of the suspicious model from the neighboring target classes by performing gradient descent optimization. 
When given a white-box access, the dataset copyright auditing methods can more accurately extract information from the model and achieve better performance. 

\mypara{Observation 5: Almost all methods require some additional auditing cost}
In the Auditing Cost column, we also show the additional auditing overhead of the existing methods in addition to access to the target dataset and the suspicious model. 
The Dataset Modification column illustrates whether the auditing method needs to manipulate the original dataset. 
The Auxiliary Dataset column indicates whether an auxiliary dataset is needed during the auditing process. 
From \autoref{tab:overview-of-existing-work}, we note that almost all strategies require a modification to the original dataset or an auxiliary dataset to perform the audit. 
There are some exceptions~\cite{DBLP:conf/icml/Choquette-ChooT21} that make the decision using a preset threshold instead of an auxiliary dataset. 

\mypara{Observation 6: Stealthiness and robustness are key considerations in the evaluation of existing work} 
Given that adversaries may introduce mechanisms to hinder auditing, \ie, adaptive attackers~\cite{Carlini2017AdversarialEA}, existing solutions typically demonstrate their defensive capabilities in two main aspects. 
Stealthiness means that watermarked samples cannot be easily detected and subsequently filtered out by attackers prior to training. 
Robustness refers to the resilience of auditing effectiveness against operations within the ML pipeline, such as dataset pre-processing, differential privacy perturbations, and model fine-tuning. 
However, existing studies evaluate the robustness of their methods at different stages of the ML pipeline, making it difficult to compare their robustness in practical settings. 
Therefore, in ~\autoref{sec:copyright-auditing-in-the-wild}, we summarize the robustness testing mechanisms used in prior work and establish five test scenarios to uniformly evaluate the robustness of different methods. 

\subsection{Taxonomy of Existing Auditing Solutions}
We note that the auditing cost can serve as a pivot to build the taxonomy of existing dataset copyright auditing methods. 
The first type needs to manipulate the original dataset, \eg, injecting backdoors~\cite{DBLP:conf/nips/LiBJYXL22, DBLP:journals/corr/abs-2010-05821, DBLP:journals/corr/abs-2303-11470} or spurious features~\cite{DBLP:journals/corr/abs-2208-13893} into the samples, and we note them as \textit{intrusive auditing}. 
The second type refers to mechanisms that do not require manipulating the original dataset. 
Instead, the majority requires an auxiliary dataset to form the basis for auditing, \ie, \textit{non-intrusive auditing} mechanisms. 

The analysis presented below mainly focuses on research in the image domain, as almost all existing work has considered this domain. 
However, it is important to note that the auditing solutions demonstrated in \autoref{sec:intrusive-auditing} and \autoref{sec:non-intrusive-auditing} are not restricted to the image domain and can be applied to other domains. 

\section{Intrusive Auditing}
\label{sec:intrusive-auditing}
The intuition behind the intrusive auditing involves embedding a covert and identifiable watermark into the dataset prior to its release. 
This watermark acts as a signature, undetectable in normal use but identifiable through specific techniques. 
If a dataset is used without authorization, especially in training an ML model, detecting this watermark provides concrete evidence of the dataset's origin and misuse. 

\subsection{Overview}
The intrusive auditing methods mainly consist of two operations, \ie, watermark injection and copyright validation. 
Existing intrusive auditing strategies can be classified into three types by the underlying techniques.  
The most widely adopted technique is the DNN backdoor-inspired watermark, including targeted backdoor with poisoned labels, targeted backdoor with clean labels, untargeted backdoor with poisoned labels, and untargeted backdoor with clean labels. 
The remaining techniques are radioactive data and style transformation. 
We will briefly introduce them as follows. 

\begin{figure}[t]
\centering
\includegraphics[width=0.9\hsize] {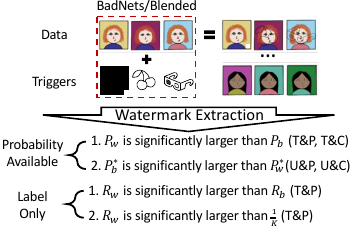}
\caption{A typical workflow of backdoor-based dataset auditing. 
The above illustrates the watermark injection process. 
These watermarks are extracted from the model's output in the model deployment phase. 
Depending on the model access permissions, the validation process can be categorized into two types: probability-based and label-only.} 
\label{fig:intrusive_auditing1}
\end{figure}

\subsection{Backdoor-based Auditing}
\label{sec:backdoor-based-auditing}

\subsubsection{Preliminaries}
\autoref{fig:intrusive_auditing1} presents a typical sequence of processes found in previous research involving backdoor injection. 
The primary distinction between these earlier works centers around the property of \textit{injected watermarks} and the corresponding \textit{copyright validation} strategies. 
Generally, in watermark injection, the owner of the dataset produces watermarked data predominantly using BadNets~\cite{GDG17} and Blended~\cite{CLLLS17} strategies. 
In copyright validation, the owner then queries the suspicious model to assess whether its dataset was used during the model training. 
If the suspicious model contains a backdoor, the dataset owner asserts that the dataset contributed to the model's training. 
Otherwise, the owner concludes that its dataset was not used in training the suspicious model. 

\subsubsection{Paper Summaries}
Existing research on backdoor-based dataset copyright auditing can be divided into four categories based on two factors. 
The first factor pertains to whether the injected backdoor is associated with a predefined label (the target label) or not, \ie, whether it is a targeted backdoor or an untargeted backdoor. 
The second factor revolves around whether the injected watermark needs to modify the true labels of the samples, \ie, whether it is a poisoned-label backdoor or a clean-label backdoor. 

\mypara{Targeted Backdoor with Poisoned Labels (T\&P)}
Li~\etal~\cite{DBLP:journals/corr/abs-2010-05821} first adopted T\&P for dataset watermarking, \ie, generating some poisoned samples by adding a local patch to some benign samples, labeled with a pre-defined target class. 
They used a hypothesis test-guided method for copyright verification. 
The copyright verification is based on the output of the suspicious model when feeding the benign samples and the corresponding watermarked samples as input. 
Then, Li~\etal~\cite{DBLP:journals/corr/abs-2209-06015, li2023black} extended the backdoor-based watermarking to other domains, such as text and graph data. 

\mypara{Targeted Backdoor with Clean Labels (T\&C)}
In contrast to T\&P, T\&C achieves the same objective by adding samples with clean labels. 
Tang~\etal~\cite{DBLP:journals/corr/abs-2303-11470} proposed to introduce imperceptible perturbations that render normal features inoperative in a few selected samples, which encourages the model to memorize the added backdoor trigger pattern. 
In the copyright validation process, the dataset owner statistically demonstrates that the addition of a secret trigger pattern can lead to changes in the prediction results, either causing them to align with the target class or significantly increasing the probability associated with the target class. 

\mypara{Untargeted Backdoor (U\&P, U\&C)} 
To address the risks associated with the T\&P and T\&C mechanisms, untargeted backdoor auditing methods refrain from specifying a target label. 
For example, Li~\etal~\cite{DBLP:conf/nips/LiBJYXL22} introduced two dispersibilities and proved their correlation, based on which they designed the untargeted backdoor watermark under both poisoned-label and clean-label settings. 
They primarily engage in dataset copyright auditing by analyzing the statistical disparities between original and watermarked samples. 

\fboxsep = 0.5mm %
\begin{center}
\begin{tcolorbox}[colback=gray!10,  %
                  colframe=black,   %
                  width=\linewidth, %
                  boxsep=-2mm,
                  breakable,
                  arc=1mm, auto outer arc,
                  boxrule=0.5pt,
                 ]
\mypara{Takeaways}
\textit{\textbf{Effectiveness vs. Stealthiness Trade-off:} 
The T\&P methods have better effectiveness, and the U\&C methods achieve better stealthiness. 
The T\&P mechanism, demonstrating a strong link with trigger-injected samples and altered labels, often outperforms other backdoor methods in efficacy. 
Furthermore, it works well both with posterior probabilities and with labels only. 
However, this approach fundamentally changes the true labels, introducing security risks and compromising the watermark's stealth. 
It enables attackers to manipulate predictions predictably using the hidden backdoor. 
Alternatively, the adversary could identify and remove these watermarked samples due to obvious label manipulation. }
\end{tcolorbox}
\end{center} 

\begin{center}
\begin{tcolorbox}[colback=gray!10,%
                  colframe=black,%
                  width=\linewidth,%
                  boxsep=-2mm,
                  breakable,
                  arc=1mm, auto outer arc,
                  boxrule=0.5pt,
                 ]
\mypara{Open Problems}
\textit{T\&C introduces potential security risks to the trained models due to the presence of poisoned labels. 
Additionally, T\&C cannot effectively handle scenarios involving label-only cases, thereby limiting its real-world applicability. 
In comparison to the T\&P and T\&C mechanisms, the U\&P and U\&C methods are characterized by their benign and covert nature. 
However, owing to the untargeted backdoors of both U\&P and U\&C, auditors may encounter challenges in correctly identifying cases of dataset theft, particularly when the performance of the suspicious model is relatively low. 
}
\end{tcolorbox}
\end{center}

\begin{figure}[t]
\centering
\includegraphics[width=0.8\hsize] {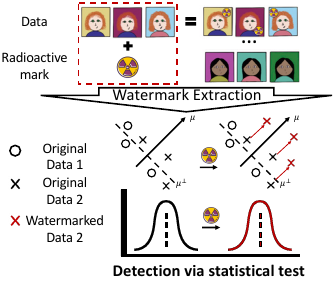}
\caption{A typical workflow of radioactive data-based auditing. 
The above illustrates the watermark injection process. 
These watermarks are extracted from the model output after training. 
The auditor determines if dataset infringement has occurred by detecting the shifts in the statistical characteristics of the model's outputs. }
\label{fig:intrusive_auditing2}
\end{figure}

\subsection{Radioactive Data-based Auditing} 
\label{sec:radioactive-data-based-auditing}
\subsubsection{Preliminaries}
From \autoref{fig:intrusive_auditing2}, radioactive data-based auditing (RDA) injects an optimized radioactive mark into the vanilla training images. 
In practice, the radioactive marks need to be propagated to the image space, which is similar to the generation of adversarial examples~\cite{GSS15, SZSBEGF14}. 
If watermarked data are used in the training, the classification model is updated with both the features and the radioactive mark. 
In copyright validation, the auditor detects the distribution deviation induced by the radioactive marks. 

\subsubsection{Paper Summaries} Sablayrolles~\etal~\cite{DBLP:conf/icml/SablayrollesDSJ20} added a random isotropic vector $\alpha u \in \mathbb{R}^d$ to the features of all vanilla training images of one class, where $\alpha$ represents the strength of the mark and $\|u\|_2=1$. 
Thus, the vector $\alpha u$ becomes the radioactive mark in the feature space. 
If there are no samples with $\alpha\mu$, $\mu$ follows a random distribution (random noise), and the cosine similarity between $w_i^T$ (fixed vector) and $\mu$ follows a beta-incomplete distribution~\cite{IFGRJ2014}. 
If the model learns samples with $\alpha\mu$, the cosine similarity between $w_i^T$ and $\mu$ will increase significantly. 
Thus, the auditing methods conduct hypothesis testing based on the above differences and decide whether the suspicious models infringe on the watermarked dataset. 
Tekgul~\etal~\cite{Tekgul2022OnTE} systematically evaluated the effectiveness of \cite{DBLP:conf/icml/SablayrollesDSJ20} on different datasets and experimental settings. 
They showed that \cite{DBLP:conf/icml/SablayrollesDSJ20} is not as effective for datasets where the number of classes is low, or the number of samples per class is low. 
Wenger~\etal~\cite{DBLP:journals/corr/abs-2208-13893} selected four kinds of marks including pixel patterns (\ie, ``pixel square'' and ``random pixels'') and blended images (\ie, ``Hello Kitty'' and ``ImageNet'' blend). 
This subset essentially instructs the model to correlate the isotope feature with the associated label. 
In addition to the above methods of superimposing radioactive marks on samples, there is also the direct injection of radioactive data into the dataset. 
Inspired by the generalization property of deep learning models, 
Guo~\etal~\cite{guo2023domain} found a hardly-generalized domain for the original dataset as the radioactive mark. 
It can be easily learned with the protected dataset containing modified samples. 
During validation, the watermarked model can correctly classify modified samples specified by the owner. 
Similarly, Li~\etal~\cite{li2023functionmarker} designed FunctionMarker, which enables LLMs to learn specific knowledge through fine-tuning on watermarked datasets and then extract the embedded watermark by obtaining the responses of LLMs to specific knowledge-related queries. 

\begin{center}
\begin{tcolorbox}[colback=gray!10,%
                  colframe=black,%
                  width=\linewidth,%
                  boxsep=-2mm,
                  arc=1mm, auto outer arc,
                  boxrule=0.5pt,
                 ]
\mypara{Takeaways}
\textit{
RDA embeds watermarks into the specified feature space by introducing marks that are orthogonal to the original dataset. 
During the validation phase, the shifts of the model's output distribution are analyzed to determine whether dataset infringement has occurred. 
}
\end{tcolorbox}
\end{center} 

\begin{center}
\begin{tcolorbox}[colback=gray!10,%
                  colframe=black,%
                  width=\linewidth,%
                  boxsep=-2mm,
                  arc=1mm, auto outer arc,
                  boxrule=0.5pt,
                 ]
\mypara{Open Problems}
\textit{
\begin{itemize}
\item Efficiency: Since the distribution shift may be slight by a single watermarked sample, the owner needs to mix a large set of marked images into the original dataset to provide statistical evidence that the model is indeed trained on watermarked images. 
\item Side effects: The radioactive mark tends to change the original dataset distribution and degrade the model's performance on normal tasks. 
\end{itemize}
}
\end{tcolorbox}
\end{center} 

\begin{figure}[t]
\centering
\includegraphics[width=0.8\hsize] {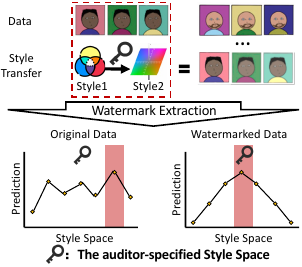}
\caption{A typical workflow of style transformation-based auditing. 
The above illustrates the watermark injection process. 
These watermarks are extracted from the model output post-training. 
The auditor conducts the audit based on the prediction of the model on the style-transferred images. }
\label{fig:intrusive_auditing3}
\end{figure}

\subsection{Style Transformation-based Auditing}
\label{sec:color-transformation-based-auditing}
\subsubsection{Preliminaries}
\autoref{fig:intrusive_auditing3} shows style transformation-based auditing (STA) uses predefined style transformations as watermark patterns for images. 
The watermark is embedded into a neural network classifier, enabling the owner to detect potential unauthorized use of its data by identifying the watermark within the neural network. 

\subsubsection{Paper Summaries} 
Zou~\etal~\cite{DBLP:conf/eccv/ZouGW22} chose to convert the original images from the RGB color space to the YIQ color space. 
During the copyright validation, the owner infers $k$ by minimizing the loss value of the suspicious model and then determines whether the suspicious model is innocent by comparing the inferred $\hat{k}$ and the true $k$. 
Li~\etal~\cite{li2022move} embedded the external features by tempering a few training samples with style transfer and then training a meta-classifier to determine whether the infringement occurs. 

\begin{center}
\begin{tcolorbox}[colback=gray!10,%
                  colframe=black,%
                  width=\linewidth,%
                  arc=1mm, auto outer arc,
                  boxsep=-2mm,
                  boxrule=0.5pt,
                 ]
\mypara{Takeaways}
\textit{The image's expansive style space ensures that STA can randomly assign unique watermarks to data from different users, \ie, allowing for more granular user-level auditing. 
Furthermore, the chosen watermark elevates the user images to a low-density latent space, which facilitates the model to memorize the watermark.} 
\end{tcolorbox}
\end{center} 

\begin{center}
\begin{tcolorbox}[colback=gray!10,%
                  colframe=black,%
                  width=\linewidth,%
                  arc=1mm, auto outer arc,
                  boxsep=-2mm,
                  boxrule=0.5pt,
                 ]
\mypara{Open Problems}
\textit{STA requires the owner to have certain knowledge about style transformation. 
If not, the image after the style transformation may be quite different from the original one. 
Thus, future work can consider introducing perceptual indicators, \eg, LPIPS~\cite{zhang2018unreasonable}, to reduce image distortion.}
\end{tcolorbox}
\end{center} 

\section{Non-intrusive Auditing}
\label{sec:non-intrusive-auditing}
The core idea of non-intrusive auditing is to leverage the dataset's inherent and unique characteristics as its fingerprint. 
These intrinsic attributes can be detected from the outputs of the models trained with the dataset. 
Identifying these fingerprints in a model substantiates the use of the dataset and supports copyright claims. 

\subsection{Overview}
The non-intrusive auditing methods mainly consist of two stages: fingerprint extraction and copyright validation. 
In the first stage, the characteristics of the suspicious model are extracted as fingerprints for following validation in the second stage. 
These methods can be organized into two types based on the fingerprints they extract. 
The most intuitive approach involves using the decision boundary as a fingerprint, in which the distance from the sample point to the boundary of different classes is used as a metric. 
The other type of fingerprint is characteristic of the model's behaviour. 
We will subsequently summarize their technical details below. 

\begin{figure}[t]
\centering
\includegraphics[width=0.9\hsize] {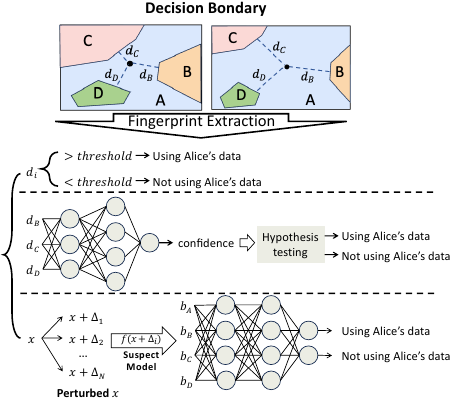}
\caption{
A typical workflow of decision boundary-based auditing. 
The existing solutions can be categorized into three types by different validation processes. 
}
\label{fig:non_intrusive_auditing1}
\end{figure}

\subsection{Decision Boundary-based Auditing}
\label{sec:decision-boundary-based-auditing}

\subsubsection{Preliminaries}
\autoref{fig:non_intrusive_auditing1} demonstrates the workflows of decision boundary-based auditing (DBA), where the decision boundary represents the ``dividing line'' between different prediction classes. 
The intuition behind DBA is that samples located on the decision boundary in the training dataset are crucial for the classification task~\cite{wang2021zero}. 
As such, the model allocates more attention to boundary samples during training to enhance classification accuracy. 
\subsubsection{Paper Summaries} 
The training data is generally far from the model's classification boundary~\cite{DBLP:conf/iclr/MainiYP21, Cao2021IPGuardPI, Karimi2019CharacterizingTD}. 
Thus, the owner can determine if a dataset is in a model training set by extracting boundary information of the model, \ie, the dataset's fingerprint. 
The methods for measuring prediction margin under white-box and black-box settings are offered in~\cite{DBLP:conf/iclr/MainiYP21, SZSBEGF14}. 
Existing solutions~\cite{DBLP:conf/icml/Choquette-ChooT21, LZ21, DBLP:conf/iclr/MainiYP21, szyller2022robustness, tian2023knowledge, DBLP:conf/icml/Choquette-ChooT21} can be divided into three major categories based on different utilization of boundary information. 

The first auditing method compares $d_i$ with a preset threshold, grounded in the principle that a classifier maximizes the distance of training examples from the decision boundary~\cite{DBLP:conf/icml/Choquette-ChooT21, LZ21}. 
The second approach involves training a network as a confidence regressor using prediction margins $\{d_i, i\in K\}$ as inputs to generate a confidence score, where $K$ is the set of all labels except the ground-truth label of the sample. 
The mean of this vector is then used in hypothesis testing to ascertain if the model utilized the target data set. 
Another method creates $N$ perturbed samples $x+\Delta_1, x+\Delta_2, \dots, x+\Delta_N$ and feeds them into the model to produce class counts $b_A, b_B, b_C, b_D$. 
These counts are fed into a deep learning model, which performs binary classification to determine the use of the target dataset~\cite{DBLP:conf/icml/Choquette-ChooT21}. 

\begin{center}
\begin{tcolorbox}[colback=gray!10,%
                  colframe=black,%
                  width=\linewidth,%
                  arc=1mm, auto outer arc,
                  boxsep=-2mm,
                  boxrule=0.5pt,
                 ]
\mypara{Takeaways}
\textit{The decision boundary mainly exists in classification tasks. 
Thus, DBA is suitable for auditing the training data of classification models. 
}
\end{tcolorbox}
\end{center} 

\begin{center}
\begin{tcolorbox}[colback=gray!10,%
                  colframe=black,%
                  width=\linewidth,%
                  arc=1mm, auto outer arc,
                  boxsep=-1.2mm,
                  boxrule=0.5pt
                 ]
\mypara{Open Problems}
\textit{The validation method based on a preset threshold is notably intuitive. 
However, it comes with significant drawbacks. 
One primary issue is the challenge of determining the appropriate threshold, which crucially influences the outcome of the judgment. 
The validation method based on hypothesis testing offers an improvement in the validation accuracy compared to the threshold-based method, while it incurs a higher computational overhead for training the regressor. 
}
\end{tcolorbox}
\end{center} 

\begin{figure}[t]
\centering
\includegraphics[width=\hsize] {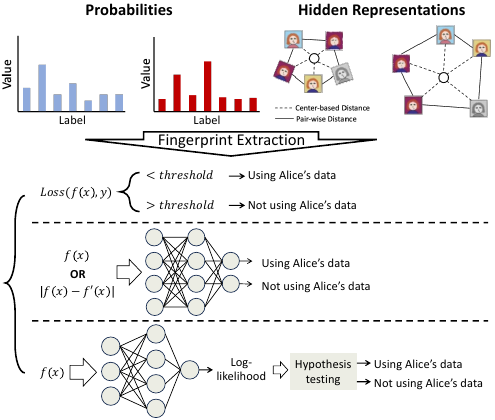}
\caption{
A typical workflow of model behavior-based auditing. 
The auditing is based on the output probabilities or the hidden representations of the model. 
Existing methods can be categorized into three types by the validation process. 
}
\label{fig:non_intrusive_auditing2}
\end{figure}

\subsection{Model Behavior-based Auditing}
\label{sec:output-characteristic-based-auditing}
\subsubsection{Preliminaries}
\autoref{fig:non_intrusive_auditing2} shows the auditing methods based on the characteristic of the model's behavior (MBA). 
In classification tasks, the model's behavior is characterized by the probabilities assigned to different labels. 
In contrast, for other tasks, the model's behavior typically involves identifying hidden representations or structures within the data, such as clusters, densities, or associations, without relying on pre-labeled responses or categories. 

\subsubsection{Paper Summaries} 
Existing solutions can be divided into three categories according to different utilizations of the model's behavior. 
The first auditing method compares the loss value, $Loss(f(x), y^{\prime})$, to a preset threshold, especially useful in classification tasks where $y^{\prime} \neq y$ indicates higher loss for the suspicious model $f$~\cite{SDSOJ19}. 
The second approach trains a discriminator using an auxiliary dataset~\cite{Xu2022DataOI, SSSS17, liu2022your, DBLP:conf/ndss/DongLCXZ023, SZHBFB19}, with inputs typically comprising the suspicious model's output $f(x)$ and the disparity between the outputs of the target and shadow models $|f(x)-f^{\prime}(x)|$. 
Another strategy estimates the log-likelihood value based on $f(x)$ and applies hypothesis testing for validation~\cite{DBLP:journals/corr/abs-2209-09024}. 
This is based on the premise that a model trained on the target dataset shows a significantly higher log-likelihood value than one without training on it. 
For instance, Li~\etal~\cite{li2022user} conducted auditing based on the compactness of the samples' hidden representations. 
The key observation to launch the validation is that the dataset owner whose data has been used during training forms more compact clusters in the latent space. 
Chen~\etal~\cite{CZWBZ23} formulated the auditing process as a user-level membership inference problem and used the similarity scores between the query image and the support set returned by the model as the basic auditing feature. 

\begin{center}
\begin{tcolorbox}[colback=gray!10,%
                  colframe=black,%
                  width=\linewidth,%
                  arc=1mm, auto outer arc,
                  boxsep=-2mm,
                  boxrule=0.5pt,
                 ]
\mypara{Takeaways}
\textit{The MBA strategies have a wider range of application scenarios than DBA. 
Since MBA utilizes the model's outputs and hidden representations as the dataset's fingerprints, it can be adapted to other tasks besides classification tasks. 
Furthermore, the suspicious models discussed are no longer limited to the supervised models and can also include unsupervised models~\cite{DBLP:journals/corr/abs-2209-09024}. }
\end{tcolorbox}
\end{center} 

\begin{center}
\begin{tcolorbox}[colback=gray!10,%
                  colframe=black,%
                  width=\linewidth,%
                  arc=1mm, auto outer arc,
                  boxsep=-1.8mm,
                  boxrule=0.5pt,
                 ]
\mypara{Open Problems}
\textit{
The MBA methods usually require the use of an auxiliary dataset to establish an auditing basis, \eg, the auditing methods~\cite{DBLP:conf/icml/Choquette-ChooT21, LZ21, SDSOJ19} can choose a more suitable threshold value by utilizing an auxiliary dataset. 
The distribution of datasets in practice may be varied compared to the test benchmarks. 
Thus, selecting a proper auxiliary dataset may be a bottleneck for MBA. 
}
\end{tcolorbox}
\end{center} 

\section{Copyright Auditing in the Wild}
\label{sec:copyright-auditing-in-the-wild}
\begin{figure*}[ht]
\centering
\includegraphics[width=\textwidth] {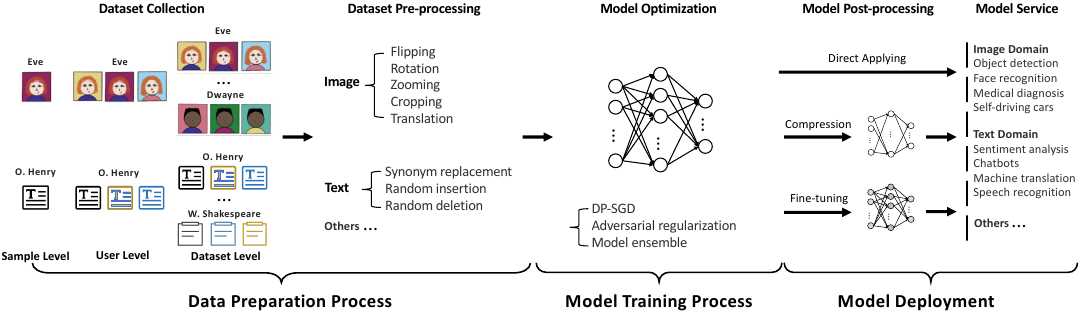}
\caption{An illustration of the machine learning system. We separate the whole workflow into three parts, \ie, the data preparation process, the model training process, and the model deployment. }
\label{fig:ml_system}
\end{figure*}
\rev{
In this section, we compare the performance of existing solutions in real-world settings. 
The evaluation has two primary objectives. 
First, we aim to compare their effectiveness under the same experimental settings, given the variations across different studies. 
Second, we evaluate how practical influencing factors affect the effectiveness of these methods, which will better inform their adoption in practice. }
To ease the understanding of the later discussions, we provide an overview of ML system's pipeline as shown in \autoref{fig:ml_system}. 
The goal is to describe the processes of the raw data from the dataset collection to the model service. 
We use it as the coordinate to locate real-world application challenges to existing solutions. 
We separate the whole workflow of \autoref{fig:ml_system} into three parts, \ie, the data preparation process (DPP), the model training process (MTP), and the model deployment (MD). 
The behavior of a suspicious model is not only determined by the owner's dataset but also incorporates aspects of DPP, MTP, and MD in ML systems. 
Thus, \autoref{sec:problem-statement} only describes an ideal situation, while \autoref{eq:dataset-auditing-problem} takes into account the impact of common data operations of practical ML systems. 
\begin{small}
\begin{equation}
\mathcal{A}: g\left(x, f_{\theta}: \textcolor{black}{P^{MD}_{\gamma}}\left(\textcolor{black}{P^{MTP}_{\beta}}\left(\textcolor{black}{P^{DPP}_{\alpha}}(x), \theta\right)\right)\right) \rightarrow \text{0 or 1}, 
\label{eq:dataset-auditing-problem}
\end{equation}
\end{small}
where $P^{DPP}_{\alpha}(x)$ represents the data preparation process, $\alpha$ denotes the pre-processing settings, and $x$ is the dataset.  
The model training process is encapsulated in $P^{MTP}_{\beta}$, involving training hyper-parameters $\beta$ and the model parameters $\theta$. 
Finally, $P^{MD}_{\eta}$ signifies the model deployment phase, with $\eta$ capturing the deployment specifics. 
The decision function $g$ integrates these stages to determine whether the adversarial model trainer has used the target dataset. 

\subsection{Experimental Setups}
Following ~\autoref{eq:dataset-auditing-problem}, we divide the ML pipeline into three parts in \autoref{fig:ml_system}, and summarize the operations that may affect the effectiveness of the auditing methods. 
To facilitate comparison, we categorize the operations into two classes. 
The first category is what model trainers normally use, which has an improvement effect on a certain attribute of the model. 
For example, model trainers usually use data augmentation to enhance the generalization ability of the model, and use differentially private stochastic gradient descent (DP-SGD)~\cite{ACGMMTZ16} to protect the privacy of training data~\cite{WIDCC19, WIDCC20}. 
The second is the adversarial mechanism proposed against the auditing method. 
Thus, the auditing scenarios can be classified into three types: ideal scenario, practical scenario, and adversarial scenario. 
The ideal scenario is that there are no interference operations during the model training process. 
The practical scenario involves the commonly used process that has an improvement effect on the model. 
The adversarial scenario pertains to some targeted anti-auditing treatment. 
In the following, we first introduce the experimental setup and then summarize the evaluation results. 

\begin{table}[t]
\scriptsize
\centering
\caption{\rev{Statistics of the \textbf{used models} in existing studies.}}
\label{tab:models-used-in-existing-paper}
\begin{tabular}{ccc}
\toprule
\textbf{Model   Name}   & \textbf{Number} & \textbf{References}                                                                                                                                                                                                                                                            \\ \midrule
ResNet-18               & 13                    & \cite{li2023black, DBLP:conf/ndss/DongLCXZ023, DBLP:journals/corr/abs-2010-05821, DBLP:conf/ijcai/HuSDCSZ22,   DBLP:conf/icml/SablayrollesDSJ20, DBLP:journals/corr/abs-2208-13893, DBLP:conf/nips/LiBJYXL22, DBLP:journals/corr/abs-2209-06015,   DBLP:journals/corr/abs-2303-11470, SDSOJ19, LZ21, Tekgul2022OnTE,   tekgul2022effectiveness} \\
ResNet-50               & 5                     & \cite{DBLP:conf/ndss/DongLCXZ023, DBLP:conf/icml/SablayrollesDSJ20, DBLP:conf/eccv/ZouGW22,   DBLP:journals/corr/abs-2208-13893, SSWM20}                                                                                                                                                                                             \\
VGG-19                  & 5                     & \cite{DBLP:conf/ndss/DongLCXZ023, li2023black, DBLP:journals/corr/abs-2010-05821,   DBLP:journals/corr/abs-2209-06015, LZ21}                                                                                                                                                                                                                      \\
LSTM                    & 4                     & \cite{DBLP:journals/corr/abs-2209-06015, SSWM20, LZ21, li2023black}                                                                                                                                                                                                                                      \\
WordCNN                 & 3                     & \cite{DBLP:journals/corr/abs-2209-06015, LZ21, li2023black}                                                                                                                                                                                                                                              \\
GIN                     & 3                     & \cite{DBLP:journals/corr/abs-2209-06015, LZ21, li2023black}                                                                                                                                                                                                                                              \\
GraphSAGE               & 3                     & \cite{DBLP:journals/corr/abs-2209-06015, LZ21, li2023black}                                                                                                                                                                                                                                              \\
CNN            & 3                     & \cite{SSSS17, SSWM20, LF20}                                                                                                                                                                                                                                                                              \\
 \bottomrule
\end{tabular}
\end{table}

\begin{table}[t]
\centering
\scriptsize
\caption{\rev{Statistics of the \textbf{used datasets} in existing studies.}}
\label{tab:datasets-used-in-existing-paper}
\begin{tabular}{ccc} 
\toprule
\textbf{Dataset Name}         & \textbf{Number} & \textbf{References}                                                                                                                                                                                                                                                                                                                                                                                                                         \\ 
\midrule
CIFAR-10              & 19     & \cite{DBLP:journals/corr/abs-2010-05821, DBLP:conf/ijcai/HuSDCSZ22, DBLP:conf/eccv/ZouGW22, DBLP:conf/nips/LiBJYXL22, DBLP:journals/corr/abs-2209-06015, DBLP:journals/corr/abs-2303-11470, DBLP:conf/iclr/MainiYP21, DBLP:journals/corr/abs-2209-09024, DBLP:conf/ndss/DongLCXZ023, SSSS17, SSWM20, DBLP:conf/icml/Choquette-ChooT21, LZ21, guo2023domain, LF20, Tekgul2022OnTE, tekgul2022effectiveness, li2023black, SDSOJ19}  \\
CIFAR-100             & 11     & \cite{DBLP:conf/eccv/ZouGW22, DBLP:journals/corr/abs-2208-13893, DBLP:conf/iclr/MainiYP21, DBLP:journals/corr/abs-2209-09024, DBLP:conf/ndss/DongLCXZ023, SSSS17, SDSOJ19, DBLP:conf/icml/Choquette-ChooT21, LF20, Tekgul2022OnTE, tekgul2022effectiveness}                                                                                                                                                                       \\
ImageNet             & 7      & \cite{DBLP:conf/icml/SablayrollesDSJ20, DBLP:conf/iclr/MainiYP21, DBLP:journals/corr/abs-2209-09024, SDSOJ19, SSWM20, DBLP:conf/icml/Choquette-ChooT21, li2023black}                                                                                                                                                                                                                                                              \\
IMDB                 & 6      & \cite{DBLP:journals/corr/abs-2209-06015, DBLP:journals/corr/abs-2303-11470, Xu2022DataOI, LZ21, Liu2023WatermarkingCD, li2023black}                                                                                                                                                                                                                                                                                               \\
Tiny ImageNet        & 4      & \cite{DBLP:conf/eccv/ZouGW22, DBLP:journals/corr/abs-2303-11470, DBLP:conf/ndss/DongLCXZ023, guo2023domain}                                                                                                                                                                                                                                                                                                                       \\
ImageNet (subset) & 3      & \cite{DBLP:conf/nips/LiBJYXL22, DBLP:journals/corr/abs-2209-06015, LZ21}                                                                                                                                                                                                                                                                                                                                                          \\
DBpedia              & 3      & \cite{DBLP:journals/corr/abs-2209-06015, LZ21, li2023black}                                                                                                                                                                                                                                                                                                                                                                       \\
COLLAB               & 3      & \cite{DBLP:journals/corr/abs-2209-06015, LZ21, li2023black}                                                                                                                                                                                                                                                                                                                                                                       \\
REDDIT-MULTI-5K      & 3      & \cite{DBLP:journals/corr/abs-2209-06015, LZ21, li2023black}                                                                                                                                                                                                                                                                                                                                                                       \\
MNIST                & 3      & \cite{SSSS17, LF20, liu2022membership}                                                                                                                                                                                                                                                                                                                                                                                            \\
\bottomrule
\end{tabular}
\end{table}

\mypara{Dataset and Model Selection}
From the statistics in \autoref{tab:models-used-in-existing-paper} and \autoref{tab:datasets-used-in-existing-paper}, CIFAR-10, CIFAR-100, ResNet, and VGG are the frequently used datasets and models in existing works. 
Thus, we utilize CIFAR-10 and CIFAR-100~\cite{krizhevsky2009learning} in the experiment. 
CIFAR-10 consists of 50000 training images and 10000 testing images divided into 10 classes. 
CIFAR-100 is structured similarly but contains 100 classes. 
Considering the misuse of facial data in the real world~\cite{Hill.2021}, we also conduct the evaluation on PubFig~\cite{kumar2009attribute}. 
We employ ResNet-18~\cite{heresnet} and VGG-19~\cite{VGG19} as target models. 
Both models are renowned for their performance in image classification tasks and are well-suited for deep learning training in complex image recognition. 

\mypara{Metrics} \textbf{B-Acc}: Model's classification accuracy on benign samples before watermarking. 
\textbf{W-Acc}: Model's classification accuracy after watermarking. 
\textbf{A-Acc}: Correct detection rate of watermark or fingerprint information. 
In practice, A-Acc is hard to achieve 100\% due to false positive and false negative cases. 
In the legal context, ``false positive'' indicates an innocent model owner incorrectly identified as an infringer, and ``false negative'' represents an infringer successfully evading the auditing. 

\mypara{Training} 
We consider six combinations of dataset and model: \textit{D1M1, D1M2, D2M1, D2M2, D3M1, D3M2}, where D1, D2, and D3 correspond to CIFAR-10, CIFAR-100, and PubFig, respectively, while M1 and M2 represent ResNet-18 and VGG-19. 
For each combination, we train the model for 100 epochs using a batch size of 32, employing cross-entropy loss as the criterion and optimizing with SGD at a learning rate of 1e-3. 

\mypara{Auditing Scenarios}
In \autoref{tab:ml-data-processing}, we summarize the data pre-processing, model optimization, and model post-processing operations used in existing work. 
We count the most commonly used operations and build two types of evaluation scenarios. 
Concretely, in practical scenarios, we implement four data augmentation techniques for dataset preprocessing, \ie, random horizontal flipping, random cropping, random cutouts, and the addition of Gaussian noise. 
These methods collectively increase the size of the dataset by a factor of five. 
On the basis of data augmentation, we utilize the differentially private stochastic gradient descent (DP-SGD) \cite{ACGMMTZ16}, a widely adopted approach in privacy-preserving contexts. 
In~\autoref{tab:evaluation-in-the-practical-scenario}, ``Prac. 1'' and ``Prac. 2'' both utilize identical dataset preprocessing methods, featuring four data augmentation types, though they differ in the noise multipliers for DP-SGD, specifically 0.8 for ``Prac. 1'' and 1.0 for ``Prac. 2''.
In adversarial scenarios, we integrate three anti-auditing strategies: fine-tuning on clean data, neural cleanse~\cite{WYSLVZZ19}, and output perturbation, referred to as ``Adv. 1'', ``Adv. 2'', and ``Adv. 3'' in~\autoref{tab:evaluation-in-the-adversarial-scenario}. 

\subsubsection{Intrusive Auditing}
We incorporate eight types of watermark setting: \textbf{BA1}(targeted label \& badnets watermark), \textbf{BA2} (clean label \& badnets watermark), \textbf{BA3} (targeted label \& blended watermark), \textbf{BA4} (clean label \& blended watermark), \textbf{U\&P} (untargeted backdoor with poisoned label), \textbf{U\&C} (untargeted backdoor with clean label), \textbf{RDA} (radioactive data-based auditing), and \textbf{STA} (style transformation-based auditing), where BA1 and BA3 belong to \textbf{T\&P} (targeted backdoor with poisoned label), BA2 and BA4 belong to \textbf{T\&C} (targeted backdoor with clean label). 
For each type of watermark, we consider three watermarking rates $\gamma$, \ie, 0.01, 0.05, and 0.1. 
Specifically, in the practical scenario, we choose two hyper-parameters of noise multiplier for DP-SGD, which are 0.8 and 1.0. 
In the adversarial scenario, we retrain the model on a benign dataset with the same training settings for another 10 epochs of fine-tuning. 
For output perturbation, we add Gaussian noise with a standard deviation of 0.01 to the normalized model output. 

\subsubsection{Non-intrusive Auditing}
We consider three types of auditing methods here: \textbf{DBA} (decision boundary-based auditing), \textbf{MBA1}, \textbf{MBA2} (model behavior-based auditing). 
In the non-intrusive auditing approach, we exclusively use CIFAR-10 for the experiments. 
We first randomly divide the training set into two parts, named CIFAR-10-A and CIFAR-10-B. 
We once select one dataset as the target dataset, and the other becomes the shadow dataset. 
Before starting the audit, we first train the two models with the target dataset and the shadow dataset, respectively. 
From each dataset, we randomly select 100 samples from each label, \ie, 1000 samples for fingerprint extraction, named CIFAR-10-S. 

For DBA, we use MinGD~\cite{DBLP:conf/iclr/MainiYP21} to extract the distance to each class boundary for the entire CIFAR-10-S dataset, obtaining $d$. 
Samples from the shadow dataset are labeled as 0, while those from the target dataset are labeled as 1. 
We use the $(d, label)$ pairs to train a simple binary classifier, which predicts whether the input fingerprints belong to the target dataset or the shadow dataset. 

For MBA1 and MBA2, we directly use the output of the target model $f$ as fingerprints. 
In MBA1, similar to DBA, we use $(f, label)$ pairs to train a binary classifier. 
For MBA2, we label the samples from the target dataset as -1 and those from the shadow dataset as 1. 
We then train a simple regressor with the $(f, label')$ pairs to predict the confidence of the input fingerprints belonging to the target dataset. 

By default, the binary classifier is a three-layer linear network with ReLU and sigmoid activations. 
The classifier's criterion is BCE loss. 
The regressor is a two-layer linear network with Tanh activations, and the loss function is $L = -g(x) \cdot label'$, where $g(x)$ refers to the output of the regressor. 
For both the binary classifier and the regressor, the optimizer is SGD with a learning rate of 5e-3. 

\begin{table}
\centering
\caption{Auditing performance evaluation in the ideal scenario. 
Without watermark injection, the classification accuracy (B-Acc) is 91.09\% for all solutions. 
Thus, we omit B-Acc in the following table to save space. 
$\gamma$ represents the proportion of watermarking samples in the entire dataset. 
}
\label{tab:evaluation-in-the-ideal-scenario-1}
\arrayrulecolor{black}
\begin{tabular}{c!{\color{black}\vrule}c!{\color{black}\vrule}c!{\color{black}\vrule}c!{\color{black}\vrule}c!{\color{black}\vrule}c!{\color{black}\vrule}c} 
\arrayrulecolor{black}\toprule
\multirow{2}{*}{\begin{tabular}[c]{@{}c@{}}\textbf{Auditing}\\\textbf{Method}\end{tabular}} & \multicolumn{2}{c!{\color{black}\vrule}}{\textbf{$\gamma = 0.01$}} & \multicolumn{2}{c!{\color{black}\vrule}}{$\gamma = 0.05$} & \multicolumn{2}{c}{$\gamma = 0.1$}  \\ 
                                         & W-Acc              & A-Acc                                  & W-Acc              & A-Acc                                  & W-Acc              & A-Acc            \\ 
\midrule
BA1                                      & 83.88              & 95.96                                  & 83.46              & 96.64                                  & 83.42              & 96.9             \\ 
BA2                                      & 84.21              & 80.77                                  & 84.16              & 82.2                                   & 83.52              & 84.5             \\ 
BA3                                      & 84.05              & 95.61                                  & 83.57              & 96.31                                  & 82.86              & 96.2             \\ 
BA4                                      & 83.98              & 82.13                                  & 83.73              & 87.62                                  & 84.15              & 87.01            \\ 
U\&P  & 91.95 & 84.47 & 91.48 & 85.23 & 90.54 & 88.43\\
U\&C & 88.78 & 83.16 & 87.72& 83.99 & 85.49 &  84.23\\
RDA                                      & 85.78              & 93.01                                  & 85.71              & 93.28                                  & 84.22              & 95.97            \\ 
STA                                      & 65.81              & 91.87                                  & 65.91              & 92.61                                  & 63.95              & 95.2             \\ 
\arrayrulecolor{black}\midrule
DBA                                      & \multirow{3}{*}{/} & 82.39                                  & \multirow{3}{*}{/} & 82.39                                  & \multirow{3}{*}{/} & 82.39            \\ 
MBA1                                     &                    & 72.27                                  &                    & 72.27                                  &                    & 72.27            \\ 
MBA2                                     &                    & 76.44                                  &                    & 76.44                                  &                    & 76.44            \\
\arrayrulecolor{black}\bottomrule
\end{tabular}
\end{table}

\begin{table*}
\centering
\caption{The performance evaluation in the ideal scenario. 
The values in the W-Acc column represent the change compared to the corresponding B-Acc. 
The values in the A-Acc column represent the auditing accuracy ($\gamma=0.1$). 
}
\label{tab:evaluation-in-the-ideal-scenario-2}
\begin{tabular}{c|cc|cc|cc|cc|cc|cc} 
\toprule
\multirow{3}{*}{\textbf{Settings}} & \multicolumn{2}{c|}{\textbf{D1M1}} & \multicolumn{2}{c|}{\textbf{D1M2}} & \multicolumn{2}{c|}{\textbf{D2M1}} & \multicolumn{2}{c|}{\textbf{D2M2}} & \multicolumn{2}{c|}{\textbf{D3M1}} & \multicolumn{2}{c}{\textbf{D3M2}}  \\ 
& \multicolumn{2}{c|}{B-Acc=91.09\%} & \multicolumn{2}{c|}{B-Acc=92.63\%} & \multicolumn{2}{c|}{B-Acc=89.47\%} & \multicolumn{2}{c|}{B-Acc=89.99\%} & \multicolumn{2}{c|}{B-Acc=90.10\%} & \multicolumn{2}{c}{B-Acc=91.94\%}  \\ 
& W-Acc & A-Acc                     & W-Acc & A-Acc                     & W-Acc & A-Acc                     & W-Acc & A-Acc                     & W-Acc & A-Acc                     & W-Acc  & A-Acc                    \\ 
\midrule
\textbf{BA1}            & -7.67 & 96.90                     & -2.46 & 97.40      & -8.47 & 93.63                     & -6.41 & 94.08                     & -3.38 & 93.27                     & -2.98 & 93.77                    \\
\textbf{BA2}             & -7.57 & 84.50                     & -2.22 & 76.45                     & -7.09 & 81.46                     & -5.21 & 83.99                     & -3.12 & 85.66                     & -2.68  & 85.79                    \\
\textbf{BA3}             & -8.23 & 96.20                     & -2.58 & 97.43                     & -8.36 & 93.74                     & -6.57 & 94.31                    & -3.31 & 92.32                     & -3.46 & 92.16                    \\
\textbf{BA4}            & -6.94 &87.01                     & -2.22 & 73.78                     & -7.46 & 82.61                     & -6.14 & 82.77                     & -3.39 & 84.12                     & -2.31 & 85.13                    \\
\textbf{U\&P}            & -0.55 & 88.43                     & -2.09 & 89.04                     & 0.44 & 86.79                     & -0.92 & 86.94                     & 0.44 & 85.12                     & -1.40 & 85.89                    \\
\textbf{U\&C}            & -5.60 & 84.23                     & -7.14 & 84.99                     & -4.61 & 81.15                     & -5.97 & 81.89                     & -4.61 & 79.83                     & -6.45 & 80.84                \\
\textbf{RDA}            & -6.87 & 95.97                     & -1.84 & 94.93                     & -6.86 & 95.01                     & -5.12 & 94.74                     & -3.02 & 94.98                     & -2.45 & 94.37                    \\
\textbf{STA}            & -27.14 & 95.20                     & -7.14 & 95.05                     & -40.10 & 94.88                     & -26.04 &95.02                     & -32.73 & 93.24                     & -27.43 & 94.07                    \\
\bottomrule
\end{tabular}
\end{table*}

\begin{table*}
\centering
\caption{The performance evaluation in the practical scenarios 
(watermarking rate $\gamma=0.1$). 
}
\label{tab:evaluation-in-the-practical-scenario}
\begin{tabular}{c|c|cc|cc|cc|cc|cc|cc} 
\toprule
\multirow{2}{*}{\textbf{Scenarios}} & \textbf{Settings} & \multicolumn{2}{c|}{\textbf{D1M1}}                                                    & \multicolumn{2}{c|}{\textbf{D1M2}}                                                    & \multicolumn{2}{c|}{\textbf{D2M1}}                                                    & \multicolumn{2}{c|}{\textbf{D2M2}}                                                    & \multicolumn{2}{c|}{\textbf{D3M1}}                                                    & \multicolumn{2}{c}{\textbf{D3M2}}                                                      \\
                                    & Methods           & W-Acc                                     & A-Acc                                     & W-Acc                                     & A-Acc                                     & W-Acc                                     & A-Acc                                     & W-Acc                                     & A-Acc                                     & W-Acc                                     & A-Acc                                     & W-Acc                                     & A-Acc                                      \\ 
\midrule
\multirow{8}{*}{\textbf{Ideal}}     & BA1               & {\cellcolor[rgb]{0.929,0.949,0.98}}83.42  & {\cellcolor[rgb]{0.376,0.557,0.784}}96.90 & {\cellcolor[rgb]{0.651,0.753,0.882}}90.17 & {\cellcolor[rgb]{0.357,0.545,0.78}}97.40  & {\cellcolor[rgb]{0.984,0.976,0.988}}81.00 & {\cellcolor[rgb]{0.51,0.651,0.831}}93.63  & {\cellcolor[rgb]{0.922,0.941,0.976}}83.58 & {\cellcolor[rgb]{0.494,0.639,0.827}}94.08 & {\cellcolor[rgb]{0.796,0.851,0.933}}86.72 & {\cellcolor[rgb]{0.525,0.663,0.839}}93.27 & {\cellcolor[rgb]{0.702,0.788,0.902}}88.96 & {\cellcolor[rgb]{0.506,0.647,0.831}}93.77  \\
                                    & BA2               & {\cellcolor[rgb]{0.925,0.945,0.98}}83.52  & {\cellcolor[rgb]{0.886,0.918,0.965}}84.50 & {\cellcolor[rgb]{0.643,0.745,0.878}}90.41 & {\cellcolor[rgb]{0.984,0.922,0.933}}76.45 & {\cellcolor[rgb]{0.973,0.976,0.996}}82.38 & {\cellcolor[rgb]{0.984,0.98,0.992}}81.46  & {\cellcolor[rgb]{0.643,0.745,0.878}}90.41 & {\cellcolor[rgb]{0.984,0.89,0.902}}73.78  & {\cellcolor[rgb]{0.784,0.843,0.929}}86.98 & {\cellcolor[rgb]{0.839,0.882,0.949}}85.66 & {\cellcolor[rgb]{0.69,0.78,0.898}}89.26   & {\cellcolor[rgb]{0.831,0.878,0.945}}85.79  \\
                                    & BA3               & {\cellcolor[rgb]{0.953,0.965,0.988}}82.86 & {\cellcolor[rgb]{0.404,0.58,0.796}}96.20  & {\cellcolor[rgb]{0.659,0.757,0.886}}90.05 & {\cellcolor[rgb]{0.353,0.541,0.776}}97.43 & {\cellcolor[rgb]{0.984,0.976,0.988}}81.11 & {\cellcolor[rgb]{0.506,0.651,0.831}}93.74 & {\cellcolor[rgb]{0.929,0.949,0.98}}83.42  & {\cellcolor[rgb]{0.482,0.631,0.824}}94.31 & {\cellcolor[rgb]{0.792,0.851,0.933}}86.79 & {\cellcolor[rgb]{0.565,0.69,0.851}}92.32  & {\cellcolor[rgb]{0.722,0.8,0.906}}88.48   & {\cellcolor[rgb]{0.573,0.694,0.855}}92.16  \\
                                    & BA4               & {\cellcolor[rgb]{0.898,0.925,0.969}}84.15 & {\cellcolor[rgb]{0.784,0.843,0.929}}87.01 & {\cellcolor[rgb]{0.643,0.745,0.878}}90.41 & {\cellcolor[rgb]{0.984,0.89,0.902}}73.78  & {\cellcolor[rgb]{0.988,0.988,1}}82.01     & {\cellcolor[rgb]{0.965,0.973,0.992}}82.61 & {\cellcolor[rgb]{0.914,0.937,0.976}}83.85 & {\cellcolor[rgb]{0.957,0.965,0.988}}82.77 & {\cellcolor[rgb]{0.796,0.851,0.933}}86.71 & {\cellcolor[rgb]{0.902,0.929,0.973}}84.12 & {\cellcolor[rgb]{0.675,0.769,0.89}}89.63  & {\cellcolor[rgb]{0.859,0.898,0.957}}85.13  \\
                                    & U\&P               & {\cellcolor[rgb]{0.639,0.741,0.878}}90.54 & {\cellcolor[rgb]{0.725,0.804,0.91}}88.43  & {\cellcolor[rgb]{0.686,0.776,0.894}}89.35 & {\cellcolor[rgb]{0.698,0.784,0.898}}89.04 & {\cellcolor[rgb]{0.722,0.8,0.906}}88.50   & {\cellcolor[rgb]{0.792,0.851,0.933}}86.79 & {\cellcolor[rgb]{0.678,0.773,0.894}}89.53 & {\cellcolor[rgb]{0.784,0.847,0.929}}86.94 & {\cellcolor[rgb]{0.737,0.812,0.914}}88.13 & {\cellcolor[rgb]{0.859,0.898,0.957}}85.12 & {\cellcolor[rgb]{0.741,0.816,0.914}}88.04 & {\cellcolor[rgb]{0.827,0.875,0.945}}85.89  \\
                                    & U\&C               & {\cellcolor[rgb]{0.843,0.886,0.949}}85.49 & {\cellcolor[rgb]{0.898,0.925,0.969}}84.23 & {\cellcolor[rgb]{0.831,0.878,0.945}}85.82 & {\cellcolor[rgb]{0.867,0.902,0.957}}84.99 & {\cellcolor[rgb]{0.878,0.914,0.965}}84.66 & {\cellcolor[rgb]{0.984,0.976,0.988}}81.15 & {\cellcolor[rgb]{0.831,0.878,0.945}}85.83 & {\cellcolor[rgb]{0.984,0.984,0.996}}81.89 & {\cellcolor[rgb]{0.914,0.937,0.976}}83.81 & {\cellcolor[rgb]{0.984,0.961,0.973}}79.83 & {\cellcolor[rgb]{0.89,0.922,0.969}}84.39  & {\cellcolor[rgb]{0.984,0.973,0.984}}80.84  \\
                                    & RDA               & {\cellcolor[rgb]{0.898,0.925,0.969}}84.22 & {\cellcolor[rgb]{0.416,0.584,0.8}}95.97   & {\cellcolor[rgb]{0.627,0.733,0.875}}90.79 & {\cellcolor[rgb]{0.459,0.616,0.816}}94.93 & {\cellcolor[rgb]{0.965,0.973,0.992}}82.61 & {\cellcolor[rgb]{0.455,0.612,0.812}}95.01 & {\cellcolor[rgb]{0.871,0.906,0.961}}84.87 & {\cellcolor[rgb]{0.467,0.62,0.816}}94.74  & {\cellcolor[rgb]{0.78,0.843,0.929}}87.08  & {\cellcolor[rgb]{0.455,0.616,0.816}}94.98 & {\cellcolor[rgb]{0.682,0.773,0.894}}89.49 & {\cellcolor[rgb]{0.482,0.631,0.824}}94.37  \\
                                    & STA               & {\cellcolor[rgb]{0.98,0.773,0.784}}63.95  & {\cellcolor[rgb]{0.447,0.608,0.812}}95.20 & {\cellcolor[rgb]{0.843,0.886,0.949}}85.49 & {\cellcolor[rgb]{0.451,0.612,0.812}}95.05 & {\cellcolor[rgb]{0.976,0.6,0.608}}49.37   & {\cellcolor[rgb]{0.459,0.616,0.816}}94.88 & {\cellcolor[rgb]{0.98,0.773,0.784}}63.95  & {\cellcolor[rgb]{0.455,0.612,0.812}}95.02 & {\cellcolor[rgb]{0.976,0.694,0.706}}57.37 & {\cellcolor[rgb]{0.525,0.663,0.839}}93.24 & {\cellcolor[rgb]{0.98,0.78,0.788}}64.51   & {\cellcolor[rgb]{0.494,0.639,0.827}}94.07  \\ 
\midrule
\multirow{8}{*}{\textbf{Prac. 1}}   & BA1               & {\cellcolor[rgb]{0.976,0.612,0.624}}50.53 & {\cellcolor[rgb]{0.565,0.69,0.851}}92.30  & {\cellcolor[rgb]{0.976,0.612,0.624}}50.53 & {\cellcolor[rgb]{0.565,0.69,0.851}}92.30  & {\cellcolor[rgb]{0.976,0.573,0.584}}47.20 & {\cellcolor[rgb]{0.631,0.737,0.875}}90.65 & {\cellcolor[rgb]{0.976,0.612,0.624}}50.53 & {\cellcolor[rgb]{0.565,0.69,0.851}}92.30  & {\cellcolor[rgb]{0.976,0.675,0.686}}55.75 & {\cellcolor[rgb]{0.62,0.729,0.871}}90.95  & {\cellcolor[rgb]{0.976,0.682,0.694}}56.39 & {\cellcolor[rgb]{0.596,0.714,0.863}}91.57  \\
                                    & BA2               & {\cellcolor[rgb]{0.976,0.612,0.624}}50.55 & {\cellcolor[rgb]{0.984,0.965,0.976}}80.02 & {\cellcolor[rgb]{0.976,0.612,0.624}}50.55 & {\cellcolor[rgb]{0.984,0.878,0.89}}73.02  & {\cellcolor[rgb]{0.976,0.639,0.647}}52.63 & {\cellcolor[rgb]{0.984,0.925,0.933}}76.69 & {\cellcolor[rgb]{0.976,0.678,0.69}}56.18  & {\cellcolor[rgb]{0.98,0.827,0.835}}68.48  & {\cellcolor[rgb]{0.976,0.694,0.706}}57.49 & {\cellcolor[rgb]{0.984,0.965,0.973}}79.97 & {\cellcolor[rgb]{0.98,0.714,0.722}}58.87  & {\cellcolor[rgb]{0.984,0.953,0.965}}79.18  \\
                                    & BA3               & {\cellcolor[rgb]{0.98,0.737,0.745}}60.90  & {\cellcolor[rgb]{0.427,0.592,0.804}}95.67 & {\cellcolor[rgb]{0.98,0.737,0.745}}60.90  & {\cellcolor[rgb]{0.427,0.592,0.804}}95.67 & {\cellcolor[rgb]{0.976,0.573,0.584}}47.26 & {\cellcolor[rgb]{0.635,0.741,0.878}}90.61 & {\cellcolor[rgb]{0.98,0.737,0.745}}60.90  & {\cellcolor[rgb]{0.427,0.592,0.804}}95.67 & {\cellcolor[rgb]{0.976,0.671,0.678}}55.34 & {\cellcolor[rgb]{0.647,0.749,0.882}}90.30 & {\cellcolor[rgb]{0.976,0.69,0.698}}56.88  & {\cellcolor[rgb]{0.655,0.757,0.886}}90.08  \\
                                    & BA4               & {\cellcolor[rgb]{0.98,0.706,0.714}}58.29  & {\cellcolor[rgb]{0.827,0.875,0.945}}85.92 & {\cellcolor[rgb]{0.98,0.706,0.714}}58.29  & {\cellcolor[rgb]{0.984,0.863,0.875}}71.59 & {\cellcolor[rgb]{0.976,0.682,0.69}}56.21  & {\cellcolor[rgb]{0.984,0.945,0.953}}78.33 & {\cellcolor[rgb]{0.98,0.706,0.714}}58.29  & {\cellcolor[rgb]{0.984,0.965,0.976}}80.08 & {\cellcolor[rgb]{0.976,0.694,0.702}}57.28 & {\cellcolor[rgb]{0.984,0.965,0.976}}80.01 & {\cellcolor[rgb]{0.98,0.71,0.722}}58.73   & {\cellcolor[rgb]{0.984,0.965,0.976}}80.06  \\
                                    & U\&P               & {\cellcolor[rgb]{0.984,0.867,0.878}}71.79 & {\cellcolor[rgb]{0.886,0.918,0.965}}84.52 & {\cellcolor[rgb]{0.98,0.835,0.847}}69.32  & {\cellcolor[rgb]{0.839,0.882,0.949}}85.65 & {\cellcolor[rgb]{0.984,0.871,0.882}}72.20 & {\cellcolor[rgb]{0.984,0.984,1}}82.11     & {\cellcolor[rgb]{0.984,0.843,0.855}}69.92 & {\cellcolor[rgb]{0.953,0.965,0.988}}82.89 & {\cellcolor[rgb]{0.984,0.867,0.878}}71.99 & {\cellcolor[rgb]{0.945,0.961,0.988}}83.01 & {\cellcolor[rgb]{0.98,0.835,0.847}}69.19  & {\cellcolor[rgb]{0.957,0.969,0.992}}82.72  \\
                                    & U\&C               & {\cellcolor[rgb]{0.98,0.839,0.847}}69.40  & {\cellcolor[rgb]{0.984,0.98,0.992}}81.36  & {\cellcolor[rgb]{0.98,0.812,0.824}}67.35  & {\cellcolor[rgb]{0.957,0.969,0.992}}82.74 & {\cellcolor[rgb]{0.98,0.839,0.847}}69.46  & {\cellcolor[rgb]{0.984,0.933,0.945}}77.36 & {\cellcolor[rgb]{0.98,0.8,0.812}}66.39    & {\cellcolor[rgb]{0.984,0.941,0.953}}78.26 & {\cellcolor[rgb]{0.98,0.812,0.824}}67.32  & {\cellcolor[rgb]{0.984,0.925,0.937}}76.74 & {\cellcolor[rgb]{0.98,0.804,0.816}}66.59  & {\cellcolor[rgb]{0.984,0.937,0.949}}77.72  \\
                                    & RDA               & {\cellcolor[rgb]{0.98,0.8,0.812}}66.36    & {\cellcolor[rgb]{0.655,0.753,0.882}}90.13 & {\cellcolor[rgb]{0.98,0.8,0.812}}66.36    & {\cellcolor[rgb]{0.565,0.69,0.851}}92.28  & {\cellcolor[rgb]{0.98,0.714,0.722}}58.84  & {\cellcolor[rgb]{0.682,0.776,0.894}}89.41 & {\cellcolor[rgb]{0.98,0.8,0.812}}66.36    & {\cellcolor[rgb]{0.584,0.702,0.859}}91.88 & {\cellcolor[rgb]{0.98,0.722,0.729}}59.50  & {\cellcolor[rgb]{0.667,0.761,0.886}}89.83 & {\cellcolor[rgb]{0.98,0.749,0.757}}61.82  & {\cellcolor[rgb]{0.639,0.745,0.878}}90.46  \\
                                    & STA               & {\cellcolor[rgb]{0.973,0.475,0.482}}38.69 & {\cellcolor[rgb]{0.667,0.761,0.886}}89.84 & {\cellcolor[rgb]{0.973,0.475,0.482}}38.69 & {\cellcolor[rgb]{0.545,0.675,0.843}}92.83 & {\cellcolor[rgb]{0.973,0.455,0.463}}37.10 & {\cellcolor[rgb]{0.761,0.827,0.922}}87.52 & {\cellcolor[rgb]{0.973,0.475,0.482}}38.69 & {\cellcolor[rgb]{0.596,0.714,0.863}}91.59 & {\cellcolor[rgb]{0.976,0.561,0.569}}45.99 & {\cellcolor[rgb]{0.816,0.867,0.941}}86.24 & {\cellcolor[rgb]{0.976,0.608,0.616}}49.95 & {\cellcolor[rgb]{0.651,0.753,0.882}}90.21  \\ 
\midrule
\multirow{8}{*}{\textbf{Prac. 2}}   & BA1               & {\cellcolor[rgb]{0.976,0.639,0.647}}52.74 & {\cellcolor[rgb]{0.416,0.584,0.8}}95.98   & {\cellcolor[rgb]{0.976,0.639,0.647}}52.74 & {\cellcolor[rgb]{0.416,0.584,0.8}}95.98   & {\cellcolor[rgb]{0.973,0.494,0.502}}40.55 & {\cellcolor[rgb]{0.604,0.718,0.867}}91.34 & {\cellcolor[rgb]{0.976,0.639,0.647}}52.74 & {\cellcolor[rgb]{0.416,0.584,0.8}}95.98   & {\cellcolor[rgb]{0.976,0.624,0.631}}51.44 & {\cellcolor[rgb]{0.616,0.725,0.871}}91.07 & {\cellcolor[rgb]{0.976,0.627,0.639}}51.78 & {\cellcolor[rgb]{0.604,0.718,0.867}}91.38  \\
                                    & BA2               & {\cellcolor[rgb]{0.976,0.612,0.624}}50.55 & {\cellcolor[rgb]{0.98,0.984,1}}82.16      & {\cellcolor[rgb]{0.976,0.612,0.624}}50.55 & {\cellcolor[rgb]{0.984,0.902,0.914}}74.88 & {\cellcolor[rgb]{0.973,0.463,0.471}}37.98 & {\cellcolor[rgb]{0.984,0.937,0.949}}77.76 & {\cellcolor[rgb]{0.976,0.678,0.69}}56.18  & {\cellcolor[rgb]{0.984,0.847,0.855}}70.08 & {\cellcolor[rgb]{0.976,0.635,0.647}}52.53 & {\cellcolor[rgb]{0.984,0.976,0.988}}81.04 & {\cellcolor[rgb]{0.98,0.702,0.71}}57.95   & {\cellcolor[rgb]{0.984,0.973,0.984}}80.89  \\
                                    & BA3               & {\cellcolor[rgb]{0.976,0.694,0.706}}57.41 & {\cellcolor[rgb]{0.42,0.588,0.8}}95.85    & {\cellcolor[rgb]{0.976,0.694,0.706}}57.41 & {\cellcolor[rgb]{0.42,0.588,0.8}}95.85    & {\cellcolor[rgb]{0.973,0.506,0.514}}41.34 & {\cellcolor[rgb]{0.627,0.733,0.875}}90.81 & {\cellcolor[rgb]{0.976,0.694,0.706}}57.41 & {\cellcolor[rgb]{0.42,0.588,0.8}}95.85    & {\cellcolor[rgb]{0.976,0.631,0.639}}52.09 & {\cellcolor[rgb]{0.624,0.733,0.875}}90.87 & {\cellcolor[rgb]{0.976,0.631,0.643}}52.16 & {\cellcolor[rgb]{0.608,0.722,0.867}}91.29  \\
                                    & BA4               & {\cellcolor[rgb]{0.976,0.682,0.69}}56.27  & {\cellcolor[rgb]{0.875,0.91,0.961}}84.75  & {\cellcolor[rgb]{0.976,0.682,0.69}}56.27  & {\cellcolor[rgb]{0.984,0.882,0.894}}73.24 & {\cellcolor[rgb]{0.976,0.655,0.663}}53.95 & {\cellcolor[rgb]{0.984,0.945,0.957}}78.38 & {\cellcolor[rgb]{0.976,0.682,0.69}}56.27  & {\cellcolor[rgb]{0.949,0.961,0.988}}82.92 & {\cellcolor[rgb]{0.976,0.627,0.639}}51.79 & {\cellcolor[rgb]{0.984,0.98,0.992}}81.34  & {\cellcolor[rgb]{0.98,0.706,0.718}}58.49  & {\cellcolor[rgb]{0.984,0.98,0.992}}81.45   \\
                                    & U\&P               & {\cellcolor[rgb]{0.984,0.878,0.886}}72.76 & {\cellcolor[rgb]{0.937,0.953,0.984}}83.28 & {\cellcolor[rgb]{0.98,0.816,0.824}}67.43  & {\cellcolor[rgb]{0.875,0.91,0.961}}84.76  & {\cellcolor[rgb]{0.98,0.839,0.851}}69.63  & {\cellcolor[rgb]{0.918,0.937,0.976}}83.72 & {\cellcolor[rgb]{0.984,0.871,0.878}}72.08 & {\cellcolor[rgb]{0.949,0.961,0.988}}82.98 & {\cellcolor[rgb]{0.984,0.855,0.867}}70.90 & {\cellcolor[rgb]{0.984,0.961,0.973}}79.91 & {\cellcolor[rgb]{0.98,0.827,0.835}}68.47  & {\cellcolor[rgb]{0.984,0.976,0.988}}81.20  \\
                                    & U\&C               & {\cellcolor[rgb]{0.98,0.808,0.82}}67.05   & {\cellcolor[rgb]{0.984,0.957,0.969}}79.34 & {\cellcolor[rgb]{0.98,0.792,0.8}}65.48    & {\cellcolor[rgb]{0.984,0.969,0.98}}80.56  & {\cellcolor[rgb]{0.98,0.784,0.796}}65.06  & {\cellcolor[rgb]{0.984,0.906,0.918}}75.20 & {\cellcolor[rgb]{0.98,0.761,0.769}}62.87  & {\cellcolor[rgb]{0.984,0.925,0.937}}76.92 & {\cellcolor[rgb]{0.98,0.753,0.765}}62.32  & {\cellcolor[rgb]{0.984,0.918,0.929}}76.31 & {\cellcolor[rgb]{0.98,0.753,0.761}}62.21  & {\cellcolor[rgb]{0.984,0.925,0.937}}76.98  \\
                                    & RDA               & {\cellcolor[rgb]{0.98,0.714,0.725}}59.16  & {\cellcolor[rgb]{0.608,0.722,0.867}}91.26 & {\cellcolor[rgb]{0.98,0.714,0.725}}59.16  & {\cellcolor[rgb]{0.502,0.647,0.831}}93.87 & {\cellcolor[rgb]{0.976,0.604,0.612}}49.75 & {\cellcolor[rgb]{0.627,0.733,0.875}}90.80 & {\cellcolor[rgb]{0.98,0.714,0.725}}59.16  & {\cellcolor[rgb]{0.561,0.686,0.851}}92.43 & {\cellcolor[rgb]{0.976,0.682,0.69}}56.21  & {\cellcolor[rgb]{0.659,0.757,0.886}}90.05 & {\cellcolor[rgb]{0.98,0.702,0.71}}57.98   & {\cellcolor[rgb]{0.588,0.706,0.859}}91.73  \\
                                    & STA               & {\cellcolor[rgb]{0.973,0.439,0.447}}36.02 & {\cellcolor[rgb]{0.651,0.753,0.882}}90.21 & {\cellcolor[rgb]{0.973,0.439,0.447}}36.02 & {\cellcolor[rgb]{0.576,0.698,0.855}}92.04 & {\cellcolor[rgb]{0.973,0.412,0.42}}33.40  & {\cellcolor[rgb]{0.725,0.804,0.91}}88.41  & {\cellcolor[rgb]{0.973,0.439,0.447}}36.02 & {\cellcolor[rgb]{0.58,0.702,0.859}}91.94  & {\cellcolor[rgb]{0.973,0.525,0.533}}43.18 & {\cellcolor[rgb]{0.788,0.847,0.929}}86.87 & {\cellcolor[rgb]{0.976,0.576,0.588}}47.53 & {\cellcolor[rgb]{0.624,0.729,0.871}}90.92  \\
\bottomrule
\end{tabular}
\end{table*}

\begin{table*}
    \centering
    \caption{The performance evaluation in the adversarial scenarios
    (watermarking rate $\gamma=0.1$). 
    }
    \label{tab:evaluation-in-the-adversarial-scenario}
    \begin{tabular}{c|c|cc|cc|cc|cc|cc|cc} 
    \toprule
    \multirow{2}{*}{\textbf{Scenarios}} & \textbf{Settings} & \multicolumn{2}{c|}{\textbf{D1M1}}                                                    & \multicolumn{2}{c|}{\textbf{D1M2}}                                                    & \multicolumn{2}{c|}{\textbf{D2M1}}                                                    & \multicolumn{2}{c|}{\textbf{D2M2}}                                                    & \multicolumn{2}{c|}{\textbf{D3M1}}                                                    & \multicolumn{2}{c}{\textbf{D3M2}}                                                      \\
                                        & Methods           & W-Acc                                     & A-Acc                                     & W-Acc                                     & A-Acc                                     & W-Acc                                     & A-Acc                                     & W-Acc                                     & A-Acc                                     & W-Acc                                     & A-Acc                                     & W-Acc                                     & A-Acc                                      \\ 
    \midrule
    \multirow{8}{*}{\textbf{Ideal}}     & BA1               & {\cellcolor[rgb]{0.984,0.98,0.992}}83.42  & {\cellcolor[rgb]{0.38,0.561,0.788}}96.9   & {\cellcolor[rgb]{0.698,0.784,0.898}}90.17 & {\cellcolor[rgb]{0.357,0.545,0.78}}97.4   & {\cellcolor[rgb]{0.984,0.949,0.961}}81    & {\cellcolor[rgb]{0.533,0.671,0.843}}93.63 & {\cellcolor[rgb]{0.984,0.98,0.992}}83.58  & {\cellcolor[rgb]{0.514,0.655,0.835}}94.08 & {\cellcolor[rgb]{0.863,0.902,0.957}}86.72 & {\cellcolor[rgb]{0.553,0.682,0.847}}93.27 & {\cellcolor[rgb]{0.757,0.827,0.922}}88.96 & {\cellcolor[rgb]{0.529,0.667,0.839}}93.77  \\
                                        & BA2               & {\cellcolor[rgb]{0.984,0.98,0.992}}83.52  & {\cellcolor[rgb]{0.969,0.976,0.996}}84.5  & {\cellcolor[rgb]{0.686,0.776,0.894}}90.41 & {\cellcolor[rgb]{0.984,0.894,0.906}}76.45 & {\cellcolor[rgb]{0.984,0.965,0.976}}82.38 & {\cellcolor[rgb]{0.984,0.957,0.969}}81.46 & {\cellcolor[rgb]{0.957,0.965,0.988}}84.78 & {\cellcolor[rgb]{0.984,0.984,0.996}}83.99 & {\cellcolor[rgb]{0.851,0.89,0.953}}86.98  & {\cellcolor[rgb]{0.914,0.937,0.976}}85.66 & {\cellcolor[rgb]{0.741,0.816,0.914}}89.26 & {\cellcolor[rgb]{0.906,0.933,0.973}}85.79  \\
                                        & BA3               & {\cellcolor[rgb]{0.984,0.973,0.984}}82.86 & {\cellcolor[rgb]{0.412,0.584,0.8}}96.2    & {\cellcolor[rgb]{0.706,0.788,0.902}}90.05 & {\cellcolor[rgb]{0.353,0.541,0.776}}97.43 & {\cellcolor[rgb]{0.984,0.949,0.961}}81.11 & {\cellcolor[rgb]{0.529,0.667,0.839}}93.74 & {\cellcolor[rgb]{0.984,0.98,0.992}}83.42  & {\cellcolor[rgb]{0.502,0.647,0.831}}94.3  & {\cellcolor[rgb]{0.859,0.898,0.957}}86.79 & {\cellcolor[rgb]{0.596,0.714,0.863}}92.32 & {\cellcolor[rgb]{0.78,0.843,0.929}}88.48  & {\cellcolor[rgb]{0.604,0.718,0.867}}92.16  \\
                                        & BA4               & {\cellcolor[rgb]{0.984,0.988,1}}84.15     & {\cellcolor[rgb]{0.851,0.89,0.953}}87.01  & {\cellcolor[rgb]{0.686,0.776,0.894}}90.41 & {\cellcolor[rgb]{0.984,0.863,0.875}}73.78 & {\cellcolor[rgb]{0.984,0.961,0.973}}82.01 & {\cellcolor[rgb]{0.984,0.969,0.98}}82.61  & {\cellcolor[rgb]{0.984,0.984,0.996}}83.85 & {\cellcolor[rgb]{0.984,0.973,0.984}}82.77 & {\cellcolor[rgb]{0.863,0.902,0.957}}86.71 & {\cellcolor[rgb]{0.988,0.988,1}}84.12     & {\cellcolor[rgb]{0.725,0.804,0.91}}89.63  & {\cellcolor[rgb]{0.937,0.953,0.984}}85.13  \\
                                        & U\&P               & {\cellcolor[rgb]{0.682,0.773,0.894}}90.54 & {\cellcolor[rgb]{0.78,0.843,0.929}}88.43  & {\cellcolor[rgb]{0.737,0.812,0.914}}89.35 & {\cellcolor[rgb]{0.753,0.824,0.918}}89.04 & {\cellcolor[rgb]{0.78,0.843,0.929}}88.5   & {\cellcolor[rgb]{0.859,0.898,0.957}}86.79 & {\cellcolor[rgb]{0.729,0.808,0.91}}89.53  & {\cellcolor[rgb]{0.851,0.894,0.953}}86.94 & {\cellcolor[rgb]{0.796,0.855,0.933}}88.13 & {\cellcolor[rgb]{0.937,0.953,0.984}}85.12 & {\cellcolor[rgb]{0.8,0.855,0.933}}88.04   & {\cellcolor[rgb]{0.902,0.929,0.973}}85.89  \\
                                        & U\&C               & {\cellcolor[rgb]{0.922,0.941,0.976}}85.49 & {\cellcolor[rgb]{0.98,0.984,1}}84.23      & {\cellcolor[rgb]{0.906,0.929,0.973}}85.82 & {\cellcolor[rgb]{0.945,0.957,0.984}}84.99 & {\cellcolor[rgb]{0.961,0.969,0.992}}84.66 & {\cellcolor[rgb]{0.984,0.953,0.965}}81.15 & {\cellcolor[rgb]{0.906,0.929,0.973}}85.83 & {\cellcolor[rgb]{0.984,0.961,0.973}}81.89 & {\cellcolor[rgb]{0.984,0.984,0.996}}83.81 & {\cellcolor[rgb]{0.984,0.937,0.945}}79.83 & {\cellcolor[rgb]{0.973,0.98,0.996}}84.39  & {\cellcolor[rgb]{0.984,0.949,0.961}}80.84  \\
                                        & RDA               & {\cellcolor[rgb]{0.98,0.984,1}}84.22      & {\cellcolor[rgb]{0.424,0.592,0.804}}95.97 & {\cellcolor[rgb]{0.671,0.765,0.89}}90.79  & {\cellcolor[rgb]{0.475,0.627,0.82}}94.93  & {\cellcolor[rgb]{0.984,0.969,0.98}}82.61  & {\cellcolor[rgb]{0.471,0.624,0.82}}95.01  & {\cellcolor[rgb]{0.969,0.976,0.996}}84.47 & {\cellcolor[rgb]{0.482,0.631,0.824}}94.74 & {\cellcolor[rgb]{0.847,0.89,0.953}}87.08  & {\cellcolor[rgb]{0.471,0.624,0.82}}94.98  & {\cellcolor[rgb]{0.733,0.808,0.91}}89.49  & {\cellcolor[rgb]{0.502,0.647,0.831}}94.37  \\
                                        & STA               & {\cellcolor[rgb]{0.98,0.745,0.753}}63.95  & {\cellcolor[rgb]{0.459,0.616,0.816}}95.2  & {\cellcolor[rgb]{0.922,0.941,0.976}}85.49 & {\cellcolor[rgb]{0.467,0.624,0.82}}95.05  & {\cellcolor[rgb]{0.976,0.569,0.576}}49.37 & {\cellcolor[rgb]{0.475,0.627,0.82}}94.88  & {\cellcolor[rgb]{0.98,0.745,0.753}}63.95  & {\cellcolor[rgb]{0.471,0.624,0.82}}95.02  & {\cellcolor[rgb]{0.976,0.667,0.675}}57.37 & {\cellcolor[rgb]{0.553,0.682,0.847}}93.24 & {\cellcolor[rgb]{0.98,0.749,0.761}}64.51  & {\cellcolor[rgb]{0.514,0.655,0.835}}94.07  \\ 
    \midrule
    \multirow{8}{*}{\textbf{Adv. 1}}    & BA1               & {\cellcolor[rgb]{0.988,0.988,1}}84.04     & {\cellcolor[rgb]{0.906,0.929,0.973}}85.82 & {\cellcolor[rgb]{0.725,0.804,0.91}}89.59  & {\cellcolor[rgb]{0.616,0.729,0.871}}91.9  & {\cellcolor[rgb]{0.984,0.957,0.969}}81.54 & {\cellcolor[rgb]{0.588,0.706,0.859}}92.53 & {\cellcolor[rgb]{0.953,0.965,0.988}}84.84 & {\cellcolor[rgb]{0.682,0.773,0.894}}90.53 & {\cellcolor[rgb]{0.984,0.914,0.925}}78.08 & {\cellcolor[rgb]{0.592,0.71,0.863}}92.43  & {\cellcolor[rgb]{0.984,0.973,0.984}}82.85 & {\cellcolor[rgb]{0.561,0.69,0.851}}93.07   \\
                                        & BA2               & {\cellcolor[rgb]{0.969,0.973,0.992}}84.52 & {\cellcolor[rgb]{0.984,0.898,0.91}}76.61  & {\cellcolor[rgb]{0.686,0.776,0.894}}90.42 & {\cellcolor[rgb]{0.984,0.867,0.878}}74.22 & {\cellcolor[rgb]{0.984,0.965,0.976}}82.31 & {\cellcolor[rgb]{0.98,0.827,0.839}}70.9   & {\cellcolor[rgb]{0.91,0.933,0.973}}85.7   & {\cellcolor[rgb]{0.98,0.839,0.847}}71.69  & {\cellcolor[rgb]{0.984,0.929,0.941}}79.38 & {\cellcolor[rgb]{0.98,0.788,0.8}}67.68    & {\cellcolor[rgb]{0.957,0.969,0.992}}84.71 & {\cellcolor[rgb]{0.98,0.8,0.812}}68.58     \\
                                        & BA3               & {\cellcolor[rgb]{0.984,0.973,0.984}}82.9  & {\cellcolor[rgb]{0.761,0.831,0.922}}88.85 & {\cellcolor[rgb]{0.718,0.8,0.906}}89.75   & {\cellcolor[rgb]{0.745,0.82,0.918}}89.2   & {\cellcolor[rgb]{0.984,0.949,0.961}}81.02 & {\cellcolor[rgb]{0.596,0.714,0.863}}92.37 & {\cellcolor[rgb]{0.941,0.957,0.984}}85.05 & {\cellcolor[rgb]{0.725,0.804,0.91}}89.62  & {\cellcolor[rgb]{0.984,0.91,0.918}}77.54  & {\cellcolor[rgb]{0.631,0.737,0.875}}91.58 & {\cellcolor[rgb]{0.984,0.976,0.988}}83.28 & {\cellcolor[rgb]{0.612,0.725,0.871}}92.02  \\
                                        & BA4               & {\cellcolor[rgb]{0.984,0.988,1}}84.14     & {\cellcolor[rgb]{0.984,0.847,0.859}}72.52 & {\cellcolor[rgb]{0.682,0.776,0.894}}90.49 & {\cellcolor[rgb]{0.984,0.859,0.871}}73.63 & {\cellcolor[rgb]{0.984,0.969,0.98}}82.6   & {\cellcolor[rgb]{0.984,0.847,0.855}}72.36 & {\cellcolor[rgb]{0.898,0.925,0.969}}85.96 & {\cellcolor[rgb]{0.98,0.824,0.835}}70.49  & {\cellcolor[rgb]{0.984,0.918,0.929}}78.32 & {\cellcolor[rgb]{0.98,0.78,0.792}}67.02   & {\cellcolor[rgb]{0.957,0.965,0.988}}84.75 & {\cellcolor[rgb]{0.98,0.804,0.816}}68.98   \\
                                        & U\&P               & {\cellcolor[rgb]{0.663,0.761,0.886}}90.91 & {\cellcolor[rgb]{0.984,0.925,0.937}}79.05 & {\cellcolor[rgb]{0.737,0.812,0.914}}89.39 & {\cellcolor[rgb]{0.984,0.976,0.988}}83.31 & {\cellcolor[rgb]{0.824,0.875,0.945}}87.54 & {\cellcolor[rgb]{0.984,0.918,0.929}}78.25 & {\cellcolor[rgb]{0.749,0.82,0.918}}89.15  & {\cellcolor[rgb]{0.984,0.941,0.953}}80.3  & {\cellcolor[rgb]{0.769,0.835,0.925}}88.72 & {\cellcolor[rgb]{0.984,0.949,0.961}}80.87 & {\cellcolor[rgb]{0.792,0.851,0.933}}88.18 & {\cellcolor[rgb]{0.984,0.949,0.957}}80.8   \\
                                        & U\&C               & {\cellcolor[rgb]{0.922,0.941,0.976}}85.46 & {\cellcolor[rgb]{0.984,0.953,0.965}}81.32 & {\cellcolor[rgb]{0.882,0.914,0.965}}86.3  & {\cellcolor[rgb]{0.984,0.973,0.984}}82.81 & {\cellcolor[rgb]{0.984,0.984,0.996}}83.75 & {\cellcolor[rgb]{0.984,0.906,0.918}}77.4  & {\cellcolor[rgb]{0.898,0.925,0.969}}85.99 & {\cellcolor[rgb]{0.984,0.91,0.922}}77.62  & {\cellcolor[rgb]{0.961,0.969,0.992}}84.68 & {\cellcolor[rgb]{0.984,0.902,0.914}}76.97 & {\cellcolor[rgb]{0.898,0.925,0.969}}86.02 & {\cellcolor[rgb]{0.984,0.922,0.933}}78.56  \\
                                        & RDA               & {\cellcolor[rgb]{0.984,0.984,0.996}}83.92 & {\cellcolor[rgb]{0.839,0.882,0.949}}87.25 & {\cellcolor[rgb]{0.675,0.769,0.89}}90.69  & {\cellcolor[rgb]{0.729,0.808,0.91}}89.52  & {\cellcolor[rgb]{0.984,0.961,0.973}}81.9  & {\cellcolor[rgb]{0.745,0.816,0.914}}89.21 & {\cellcolor[rgb]{0.863,0.902,0.957}}86.72 & {\cellcolor[rgb]{0.753,0.824,0.918}}89.01 & {\cellcolor[rgb]{0.984,0.933,0.945}}79.67 & {\cellcolor[rgb]{0.561,0.686,0.851}}93.09 & {\cellcolor[rgb]{0.945,0.957,0.984}}85.03 & {\cellcolor[rgb]{0.608,0.722,0.867}}92.11  \\
                                        & STA               & {\cellcolor[rgb]{0.984,0.965,0.976}}82.12 & {\cellcolor[rgb]{0.984,0.929,0.941}}79.39 & {\cellcolor[rgb]{0.78,0.843,0.929}}88.48  & {\cellcolor[rgb]{0.984,0.953,0.965}}81.42 & {\cellcolor[rgb]{0.984,0.929,0.937}}79.19 & {\cellcolor[rgb]{0.984,0.949,0.961}}80.87 & {\cellcolor[rgb]{0.984,0.965,0.976}}82.19 & {\cellcolor[rgb]{0.984,0.961,0.973}}81.9  & {\cellcolor[rgb]{0.973,0.537,0.545}}46.66 & {\cellcolor[rgb]{0.918,0.937,0.976}}85.59 & {\cellcolor[rgb]{0.976,0.659,0.667}}56.78 & {\cellcolor[rgb]{0.875,0.91,0.961}}86.49   \\ 
    \midrule
    \multirow{8}{*}{\textbf{Adv. 2}}    & BA1               & {\cellcolor[rgb]{0.984,0.976,0.988}}83.15 & {\cellcolor[rgb]{0.867,0.902,0.957}}86.63 & {\cellcolor[rgb]{0.714,0.796,0.906}}89.88 & {\cellcolor[rgb]{0.839,0.886,0.949}}87.2  & {\cellcolor[rgb]{0.984,0.937,0.949}}80.1  & {\cellcolor[rgb]{0.757,0.827,0.922}}88.94 & {\cellcolor[rgb]{0.871,0.906,0.961}}86.58 & {\cellcolor[rgb]{0.988,0.988,1}}84.04     & {\cellcolor[rgb]{0.984,0.898,0.906}}76.6  & {\cellcolor[rgb]{0.855,0.894,0.953}}86.88 & {\cellcolor[rgb]{0.984,0.969,0.98}}82.69  & {\cellcolor[rgb]{0.843,0.886,0.949}}87.14  \\
                                        & BA2               & {\cellcolor[rgb]{0.984,0.98,0.992}}83.52  & {\cellcolor[rgb]{0.984,0.859,0.871}}73.45 & {\cellcolor[rgb]{0.686,0.776,0.894}}90.41 & {\cellcolor[rgb]{0.984,0.867,0.878}}74.17 & {\cellcolor[rgb]{0.984,0.969,0.98}}82.54  & {\cellcolor[rgb]{0.984,0.843,0.855}}72.13 & {\cellcolor[rgb]{0.839,0.882,0.949}}87.26 & {\cellcolor[rgb]{0.984,0.867,0.875}}73.99 & {\cellcolor[rgb]{0.984,0.929,0.941}}79.45 & {\cellcolor[rgb]{0.98,0.804,0.812}}68.78  & {\cellcolor[rgb]{0.91,0.933,0.973}}85.7   & {\cellcolor[rgb]{0.98,0.824,0.831}}70.39   \\
                                        & BA3               & {\cellcolor[rgb]{0.984,0.969,0.98}}82.52  & {\cellcolor[rgb]{0.949,0.961,0.988}}84.9  & {\cellcolor[rgb]{0.718,0.796,0.906}}89.81 & {\cellcolor[rgb]{0.847,0.89,0.953}}87.05  & {\cellcolor[rgb]{0.984,0.937,0.949}}79.92 & {\cellcolor[rgb]{0.722,0.8,0.906}}89.74   & {\cellcolor[rgb]{0.878,0.914,0.965}}86.37 & {\cellcolor[rgb]{0.98,0.984,1}}84.24      & {\cellcolor[rgb]{0.984,0.894,0.906}}76.35 & {\cellcolor[rgb]{0.831,0.878,0.945}}87.39 & {\cellcolor[rgb]{0.984,0.957,0.969}}81.56 & {\cellcolor[rgb]{0.831,0.878,0.945}}87.36  \\
                                        & BA4               & {\cellcolor[rgb]{0.984,0.984,0.996}}83.9  & {\cellcolor[rgb]{0.98,0.824,0.835}}70.63  & {\cellcolor[rgb]{0.686,0.776,0.894}}90.41 & {\cellcolor[rgb]{0.984,0.859,0.871}}73.61 & {\cellcolor[rgb]{0.984,0.961,0.973}}82.01 & {\cellcolor[rgb]{0.98,0.835,0.847}}71.5   & {\cellcolor[rgb]{0.812,0.867,0.941}}87.76 & {\cellcolor[rgb]{0.984,0.851,0.863}}72.77 & {\cellcolor[rgb]{0.984,0.937,0.949}}79.9  & {\cellcolor[rgb]{0.98,0.784,0.796}}67.36  & {\cellcolor[rgb]{0.878,0.914,0.965}}86.37 & {\cellcolor[rgb]{0.98,0.808,0.816}}69.1    \\
                                        & U\&P               & {\cellcolor[rgb]{0.804,0.859,0.937}}87.97 & {\cellcolor[rgb]{0.984,0.894,0.906}}76.51 & {\cellcolor[rgb]{0.953,0.965,0.988}}84.81 & {\cellcolor[rgb]{0.984,0.945,0.957}}80.51 & {\cellcolor[rgb]{0.984,0.973,0.984}}82.82 & {\cellcolor[rgb]{0.984,0.871,0.882}}74.52 & {\cellcolor[rgb]{0.894,0.925,0.969}}86.03 & {\cellcolor[rgb]{0.984,0.855,0.863}}73.01 & {\cellcolor[rgb]{0.949,0.961,0.988}}84.94 & {\cellcolor[rgb]{0.984,0.922,0.933}}78.57 & {\cellcolor[rgb]{0.945,0.961,0.988}}84.96 & {\cellcolor[rgb]{0.984,0.929,0.937}}79.18  \\
                                        & U\&C               & {\cellcolor[rgb]{0.984,0.984,0.996}}83.88 & {\cellcolor[rgb]{0.984,0.843,0.855}}72.14 & {\cellcolor[rgb]{0.984,0.976,0.988}}83.13 & {\cellcolor[rgb]{0.984,0.91,0.922}}77.85  & {\cellcolor[rgb]{0.984,0.965,0.976}}82.11 & {\cellcolor[rgb]{0.98,0.812,0.82}}69.42   & {\cellcolor[rgb]{0.984,0.965,0.976}}82.11 & {\cellcolor[rgb]{0.98,0.831,0.843}}71.3   & {\cellcolor[rgb]{0.984,0.957,0.969}}81.7  & {\cellcolor[rgb]{0.98,0.839,0.847}}71.76  & {\cellcolor[rgb]{0.984,0.953,0.965}}81.21 & {\cellcolor[rgb]{0.984,0.859,0.871}}73.41  \\
                                        & RDA               & {\cellcolor[rgb]{0.984,0.984,0.996}}83.82 & {\cellcolor[rgb]{0.718,0.796,0.906}}89.8  & {\cellcolor[rgb]{0.98,0.702,0.714}}60.59  & {\cellcolor[rgb]{0.694,0.784,0.898}}90.24 & {\cellcolor[rgb]{0.984,0.969,0.98}}82.61  & {\cellcolor[rgb]{0.698,0.784,0.898}}90.16 & {\cellcolor[rgb]{0.863,0.902,0.957}}86.71 & {\cellcolor[rgb]{0.686,0.776,0.894}}90.44 & {\cellcolor[rgb]{0.984,0.929,0.941}}79.41 & {\cellcolor[rgb]{0.522,0.659,0.835}}93.92 & {\cellcolor[rgb]{0.953,0.965,0.988}}84.79 & {\cellcolor[rgb]{0.525,0.663,0.839}}93.82  \\
                                        & STA               & {\cellcolor[rgb]{0.98,0.745,0.753}}63.95  & {\cellcolor[rgb]{0.984,0.941,0.953}}80.26 & {\cellcolor[rgb]{0.914,0.933,0.973}}85.69 & {\cellcolor[rgb]{0.984,0.984,0.996}}83.76 & {\cellcolor[rgb]{0.976,0.565,0.576}}49.17 & {\cellcolor[rgb]{0.984,0.953,0.965}}81.39 & {\cellcolor[rgb]{0.98,0.808,0.82}}69.36   & {\cellcolor[rgb]{0.965,0.973,0.992}}84.61 & {\cellcolor[rgb]{0.973,0.424,0.431}}37.48 & {\cellcolor[rgb]{0.886,0.918,0.965}}86.21 & {\cellcolor[rgb]{0.973,0.522,0.529}}45.42 & {\cellcolor[rgb]{0.78,0.843,0.929}}88.46   \\ 
    \midrule
    \multirow{8}{*}{\textbf{Adv. 3}}    & BA1               & {\cellcolor[rgb]{0.984,0.976,0.988}}83.09 & {\cellcolor[rgb]{0.494,0.639,0.827}}94.52 & {\cellcolor[rgb]{0.706,0.788,0.902}}90.06 & {\cellcolor[rgb]{0.384,0.565,0.788}}96.77 & {\cellcolor[rgb]{0.984,0.89,0.898}}75.96  & {\cellcolor[rgb]{0.6,0.718,0.867}}92.25   & {\cellcolor[rgb]{0.984,0.894,0.906}}76.54 & {\cellcolor[rgb]{0.553,0.682,0.847}}93.27 & {\cellcolor[rgb]{0.984,0.859,0.871}}73.63 & {\cellcolor[rgb]{0.678,0.769,0.89}}90.64  & {\cellcolor[rgb]{0.984,0.886,0.894}}75.62 & {\cellcolor[rgb]{0.788,0.847,0.929}}88.3   \\
                                        & BA2               & {\cellcolor[rgb]{0.984,0.984,1}}84.19     & {\cellcolor[rgb]{0.976,0.647,0.655}}55.78 & {\cellcolor[rgb]{0.69,0.78,0.898}}90.37   & {\cellcolor[rgb]{0.976,0.58,0.588}}50.32  & {\cellcolor[rgb]{0.984,0.906,0.918}}77.44 & {\cellcolor[rgb]{0.976,0.569,0.576}}49.36 & {\cellcolor[rgb]{0.984,0.91,0.922}}77.8   & {\cellcolor[rgb]{0.976,0.647,0.655}}55.81 & {\cellcolor[rgb]{0.984,0.871,0.882}}74.52 & {\cellcolor[rgb]{0.976,0.588,0.6}}51.1    & {\cellcolor[rgb]{0.984,0.922,0.929}}78.51 & {\cellcolor[rgb]{0.976,0.576,0.584}}50.06  \\
                                        & BA3               & {\cellcolor[rgb]{0.984,0.969,0.98}}82.72  & {\cellcolor[rgb]{0.447,0.608,0.812}}95.49 & {\cellcolor[rgb]{0.714,0.796,0.906}}89.87 & {\cellcolor[rgb]{0.443,0.604,0.808}}95.56 & {\cellcolor[rgb]{0.984,0.89,0.898}}75.91  & {\cellcolor[rgb]{0.604,0.718,0.867}}92.17 & {\cellcolor[rgb]{0.984,0.894,0.906}}76.29 & {\cellcolor[rgb]{0.494,0.639,0.827}}94.51 & {\cellcolor[rgb]{0.984,0.867,0.878}}74.27 & {\cellcolor[rgb]{0.706,0.788,0.902}}90.06 & {\cellcolor[rgb]{0.984,0.89,0.898}}75.91  & {\cellcolor[rgb]{0.804,0.859,0.937}}87.96  \\
                                        & BA4               & {\cellcolor[rgb]{0.984,0.984,0.996}}83.87 & {\cellcolor[rgb]{0.976,0.62,0.627}}53.51  & {\cellcolor[rgb]{0.678,0.769,0.89}}90.62  & {\cellcolor[rgb]{0.976,0.58,0.588}}50.26  & {\cellcolor[rgb]{0.984,0.894,0.906}}76.53 & {\cellcolor[rgb]{0.976,0.565,0.576}}49.23 & {\cellcolor[rgb]{0.984,0.898,0.91}}76.78  & {\cellcolor[rgb]{0.976,0.643,0.655}}55.69 & {\cellcolor[rgb]{0.984,0.863,0.875}}73.93 & {\cellcolor[rgb]{0.984,0.886,0.898}}75.68 & {\cellcolor[rgb]{0.984,0.929,0.941}}79.4  & {\cellcolor[rgb]{0.976,0.6,0.608}}51.84    \\
                                        & U\&P               & {\cellcolor[rgb]{0.733,0.812,0.914}}89.42 & {\cellcolor[rgb]{0.984,0.949,0.961}}80.92 & {\cellcolor[rgb]{0.784,0.847,0.929}}88.37 & {\cellcolor[rgb]{0.984,0.969,0.98}}82.71  & {\cellcolor[rgb]{0.988,0.988,1}}84.08     & {\cellcolor[rgb]{0.984,0.886,0.898}}75.85 & {\cellcolor[rgb]{0.906,0.933,0.973}}85.79 & {\cellcolor[rgb]{0.984,0.941,0.953}}80.28 & {\cellcolor[rgb]{0.984,0.969,0.98}}82.54  & {\cellcolor[rgb]{0.984,0.945,0.957}}80.7  & {\cellcolor[rgb]{0.984,0.98,0.992}}83.42  & {\cellcolor[rgb]{0.984,0.937,0.949}}80.08  \\
                                        & U\&C               & {\cellcolor[rgb]{0.984,0.973,0.984}}82.78 & {\cellcolor[rgb]{0.984,0.875,0.886}}74.86 & {\cellcolor[rgb]{0.965,0.973,0.992}}84.6  & {\cellcolor[rgb]{0.984,0.925,0.933}}78.86 & {\cellcolor[rgb]{0.984,0.929,0.941}}79.43 & {\cellcolor[rgb]{0.984,0.855,0.863}}72.99 & {\cellcolor[rgb]{0.984,0.945,0.957}}80.77 & {\cellcolor[rgb]{0.984,0.882,0.894}}75.41 & {\cellcolor[rgb]{0.984,0.941,0.953}}80.22 & {\cellcolor[rgb]{0.984,0.859,0.867}}73.31 & {\cellcolor[rgb]{0.984,0.933,0.945}}79.59 & {\cellcolor[rgb]{0.984,0.878,0.886}}74.97  \\
                                        & RDA               & {\cellcolor[rgb]{0.984,0.984,0.996}}83.93 & {\cellcolor[rgb]{0.816,0.867,0.941}}87.69 & {\cellcolor[rgb]{0.675,0.769,0.89}}90.71  & {\cellcolor[rgb]{0.698,0.784,0.898}}90.22 & {\cellcolor[rgb]{0.984,0.91,0.918}}77.57  & {\cellcolor[rgb]{0.745,0.816,0.914}}89.23 & {\cellcolor[rgb]{0.984,0.898,0.91}}76.81  & {\cellcolor[rgb]{0.651,0.749,0.882}}91.22 & {\cellcolor[rgb]{0.984,0.886,0.898}}75.86 & {\cellcolor[rgb]{0.569,0.694,0.855}}92.94 & {\cellcolor[rgb]{0.886,0.918,0.965}}86.26 & {\cellcolor[rgb]{0.58,0.702,0.859}}92.65   \\
                                        & STA               & {\cellcolor[rgb]{0.98,0.757,0.769}}65.08  & {\cellcolor[rgb]{0.718,0.796,0.906}}89.82 & {\cellcolor[rgb]{0.906,0.929,0.973}}85.85 & {\cellcolor[rgb]{0.988,0.988,1}}84.09     & {\cellcolor[rgb]{0.973,0.541,0.553}}47.25 & {\cellcolor[rgb]{0.71,0.792,0.902}}89.97  & {\cellcolor[rgb]{0.976,0.694,0.706}}59.92 & {\cellcolor[rgb]{0.925,0.945,0.98}}85.43  & {\cellcolor[rgb]{0.973,0.412,0.42}}36.22  & {\cellcolor[rgb]{0.765,0.831,0.922}}88.77 & {\cellcolor[rgb]{0.973,0.486,0.498}}42.72 & {\cellcolor[rgb]{0.757,0.827,0.922}}88.97  \\
    \bottomrule
\end{tabular}
\end{table*}    

\subsection{Highlighted Conclusions}
\label{sec:highlighted-conclusions}
From the results in \autoref{tab:evaluation-in-the-ideal-scenario-1}, \autoref{tab:evaluation-in-the-ideal-scenario-2} (ideal scenario), \autoref{tab:evaluation-in-the-practical-scenario} (practical scenario), and \autoref{tab:evaluation-in-the-adversarial-scenario} (adversarial scenario), we conclude the following observations for better adoption of the existing solutions. 

\mypara{Observations in Ideal Scenario}
\textit{Intrusive methods obtain higher auditing effectiveness than non-intrusive methods.} 
From \autoref{tab:evaluation-in-the-ideal-scenario-1}, the intrusive auditing method can achieve higher auditing accuracy, up to $96.9\%$, while the non-intrusive method can only achieve $82.39\%$ under the same settings. 
The intrusive method introduces watermarks to build an additional connection between watermarks and specific model behaviors during model training. 
This relationship is generally independent of the normal data features, so it can be easily extracted from the model behavior during auditing. 
This can also be supported by the fact that the watermarking rate has no significant impact on the intrusion method. 
Non-intrusive methods rely on the model's behavioral differences between training data and non-training data to make judgments. 
In other cases, such as when the model performance is poor, the auditing performance of non-intrusive methods often decreases significantly. 

\textit{Intrusive methods tend to negatively impact the performance of the model. }
From \autoref{tab:evaluation-in-the-ideal-scenario-2}, compared with the non-intrusive methods, the watermarked models exhibit varying degrees of performance degradation. 
Among them, the W-Acc of the STA has the highest attenuation, reaching $32.73\%$. 
Since STA changes the feature space of the original data, such as converting the original images from the RGB color space to the YIQ color space~\cite{DBLP:conf/eccv/ZouGW22}, it interferes with the model's ability to judge normal samples. 

\mypara{Observations in Practical Scenario}
\textit{Data Augmentation and DP-SGD have little effect on the intrusive methods}
From~\autoref{tab:evaluation-in-the-practical-scenario}, we find that commonly used data augmentations and DP-SGD have little effect on intrusive methods. 
For example, when data augmentation and DP-SGD are used during model training, the auditing performance of the intrusive method is attenuated by up to $5.6\%$ (RDA \& D2M1). 
However, the normal performance of the model is attenuated by $32.86\%$. 

\textit{The targeted backdoor-based auditing methods are more robust to these operations. }
The A-Acc results of targeted backdoor-based auditing methods, \ie, BA1, BA2, BA3, and BA4, are better than other auditing methods. 
For instance, BA3's A-Acc achieves $97.43\%$ in the ideal case of D1M2 and only drops at most $1.76\%$ in the practical case. 
This is partly due to the fixed patterns of the backdoor used by these auditing methods, leading the model to memorize these patterns deeply during the training process. 
Additionally, the design of these methods incorporates considerations for robustness against various operations. 
Although a targeted backdoor may bring additional risks, the related auditing methods are still the most effective. 
Thus, future work can consider how to mitigate this additional risk.  

\mypara{Observations in Adversarial Scenarios} 
\textit{BA1, BA3, and RDA show strong robustness against adversarial perturbation.} 
In \autoref{tab:evaluation-in-the-adversarial-scenario}, we mainly evaluate three adversarial strategies, \ie, fine-tuning on clean data, neural cleanse~\cite{WYSLVZZ19}, and output perturbation. 
Under the experimental setting, BA1, BA3, and RDA have almost no obvious auditing performance degradation across the adversarial settings, with a maximum of $11.30\%$. 
However, BA4 showed a relatively obvious performance degradation in all three adversarial settings, with a maximum degradation of $40.94\%$. 

\textit{For clean label backdoor-based methods, output perturbation is an efficient adversarial mechanism}. 
It can be seen from the BA2 and BA4 columns that compared with the first two adversarial methods, output perturbation can make BA2 and BA4 achieve greater performance degradation. 
Especially for D1M1, all adversarial strategies have a comparable impact on BA4's W-Acc. 
However, the reduction in A-Acc caused by output perturbation is nearly twice as large as that induced by the other two adversarial methods. 

\mypara{Guideline for Method Selection}
Recalling the existing solutions' technical details (\autoref{sec:intrusive-auditing} and \autoref{sec:non-intrusive-auditing}) and evaluation results (\autoref{sec:copyright-auditing-in-the-wild}), we create a tree diagram to assist practitioners in selecting an appropriate auditing method. 
Each node in the tree represents a different selection criterion, which is mainly determined by the auditing strategies' assumptions or best use cases. 

\begin{figure}[!t]
\centering
\includegraphics[width=0.9\hsize]{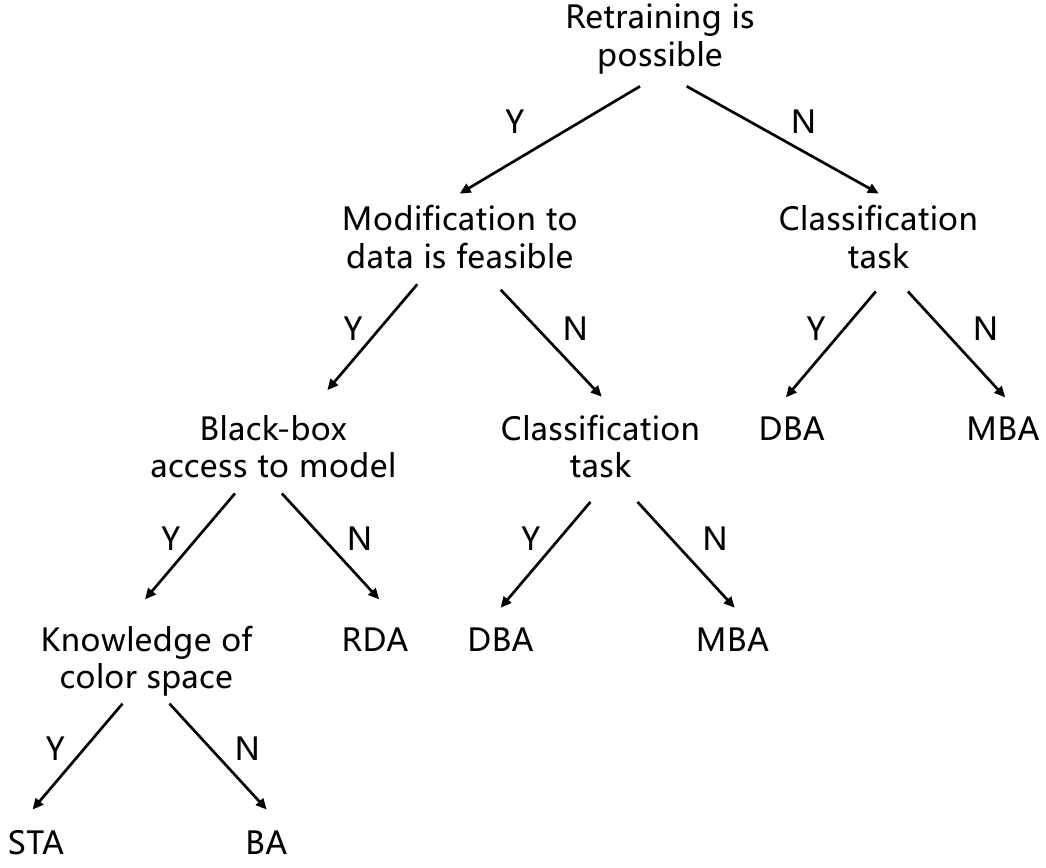}
\caption{\rev{A quick guide to select a proper auditing strategy. } 
}
\label{fig:auditing_choice}
\end{figure}

In \autoref{fig:auditing_choice}, we observe that the application of intrusive auditing methods necessitates two conditions: first, the feasibility of retraining, and second, the permission to modify the original dataset. 
If either of these conditions is not met, it becomes necessary to select between DBA and MBA based on the specific task requirements. 
Among the intrusive auditing methods, BA demonstrates strong performance under minimal requirements, making it the preferred approach. 
In the case of RDA, if the owner opts to introduce radioactive data into the feature space, internal model information, such as structure or weights, is typically required to ensure the effectiveness of the auditing. 
For STA, limited knowledge of color transformation on the part of the owner can result in noticeable color distortion in the watermarked image, potentially degrading the model's normal performance and making the watermark easily detectable by adversaries. 

\section{Promising Directions for Future Research}
\label{sec:promising-directions-in-future}
\mypara{Direction 1: Comprehensive frameworks to evaluate the impact of data preparation, model training, and deployment processes to ensure the effectiveness of dataset copyright auditing methods in actual deployment}
From \autoref{tab:ml-data-processing}, we observe that the vast majority of studies consider evaluating the robustness of their approaches in terms of data preparation, model training, and deployment processes. 
During the data preparation process, Zou~\etal~\cite{DBLP:conf/eccv/ZouGW22} and Guo~\etal~\cite{guo2023domain} utilize data augmentation as part of their evaluation. 
When it comes to the model training process, popular methods include differential privacy~\cite{LF20, DBLP:conf/icml/Choquette-ChooT21, DBLP:conf/ijcai/HuSDCSZ22, CZWBZ23, ZWLHC18, DZBLJCC21, ZWHLBHCZ21, WZWHBCZ23, YZDCCS23, WZZCMLFC23, Wang2024DPAdapterID}, dropout~\cite{SZHBFB19, LF20}, regularization~\cite{SSSS17, DBLP:conf/icml/Choquette-ChooT21, SM21}, and ensemble learning~\cite{SZHBFB19}. 
For model post-processing, existing works consider fine-tuning~\cite{DBLP:journals/corr/abs-2208-13893, DBLP:conf/iclr/MainiYP21, DBLP:journals/corr/abs-2209-06015, guo2023domain}, model pruning~\cite{DBLP:conf/iclr/MainiYP21, DBLP:journals/corr/abs-2209-06015, guo2023domain}, output perturbation~\cite{CZWBZ23, LZ21}, reducing granularity of outputs~\cite{DBLP:conf/icml/Choquette-ChooT21, DBLP:journals/corr/abs-2208-13893, DBLP:conf/ndss/DongLCXZ023} and neural cleanse~\cite{SSWM20, DBLP:conf/ijcai/HuSDCSZ22, guo2023domain}. 
It should be noted that the small amount of work that simultaneously considers the impact of these three processes on the proposed methods~\cite{CZWBZ23, DBLP:conf/ndss/DongLCXZ023, guo2023domain}. 
Thus, the future direction is to develop a comprehensive toolbox for assessing auditing effectiveness in both practical and adversarial contexts. 
The challenges originate from two sources: 1) the evaluation results of the toolbox should be close to real application scenarios; 2) Given the rapid development of models and auditing techniques, the toolbox needs to be flexible and extensible. 
One promising development idea is based on red and blue teams' confrontation, similar to Adversarial Robustness Toolbox (ART)\footnote{https://github.com/Trusted-AI/adversarial-robustness-toolbox}. 
The toolbox contains the auditing methods and robustness testing methods mentioned in the paper, as well as models and datasets to quickly build evaluation cases. 

\mypara{Direction 2: Dataset copyright auditing tools for large language models and multi-modal models}
The training data used in large language models (LLMs) has raised significant copyright concerns, which are becoming increasingly prominent as these models become more advanced and widespread~\cite{chu2023protect, sag2023copyright, peng2023you, yao2023promptcare, li2024digger}. 
For example, studies have shown that popular works are more likely to be memorized verbatim by models, which could lead to copyright violations~\cite{Karamolegkou2023CopyrightVA, casper2024black}. 
To this end, several research efforts have attempted to propose copyright protection schemes for LLMs. 
For the prompt of LLMs, Yao~\etal~\cite{yao2023promptcare} introduced PromptCARE, a watermark injection and verification scheme specifically tailored for prompts in the natural language domain. 
The framework is designed to address the challenges of watermarking in prompts, which is essential due to the growing importance of prompts in LLM-based services and the potential for their unauthorized use. 
Li~\etal~\cite{li2024digger} proposed Digger, a framework designed to identify if specific target materials were used in the training of LLM. 
There are still many issues that need attention, such as copyright protection for multi-modal data. 
We identify two main challenges. 
Firstly, the large-scale data used during the pre-training of extensive models, combined with multiple data sources, tends to dilute the effectiveness of the auditing strategies. 
As \autoref{tab:evaluation-in-the-ideal-scenario-1} illustrates, a decrease in the percentage of watermark data correlates with a lower detection success rate. 
A promising technique is domain watermarking, \eg, ~\cite{guo2023domain, li2023functionmarker}, which utilizes difficult samples within the dataset as distinct features. 
This offers two benefits: 1) the model is more likely to memorize these samples during training, aiding detection; 
2) The difficult samples are also critical for normal tasks, so filtering them to avoid auditing tends to reduce model performance on the normal tasks, \ie, elevating the cost for infringers to circumvent auditing. 
The second challenge concerns the limitations of existing methods, which are typically designed for single modalities, whereas large models are increasingly multi-modal. 
For instance, an infringer might alter the original caption of an image to bypass detection methods that focus on image content. 
Prompt optimization, such as~\cite{kepel2024autonomous}, offers a promising solution by refining prompts to better resonate with image features, thus enhancing auditing efficacy. 

\mypara{Direction 3: Dataset copyright auditing methods with formal guaranteed verification}
Current methods for dataset copyright auditing typically yield probabilistic results. 
When evaluating the effectiveness of these methods, performance is often measured using accuracy-based metrics. 
Only a handful of methods offer formal assurances regarding the reliability of their audit outcomes. 
This aspect is particularly vital for auditing tools, as the results of the audit could serve as evidence in legal actions against the owners of the model under suspicion. 
The challenges come from the inherent non-linearity of the model, the stochasticity in training, and the diversity of the distribution of the dataset. 
Traditional methods are mainly based on predefined thresholds or training a DNN-based classifier or regressor for determination. 
These methods are less interpretable in their determination and are prone to misclassification when the distribution of the auxiliary dataset and the actual auditing dataset differ significantly. 
Currently, hypothesis testing is the prevalent approach~\cite{guo2023domain, DBLP:journals/corr/abs-2010-05821, li2023black}, providing auditing results with associated significance indicators. 
Furthermore, it is suggested that future work could give accuracy at different significance levels, which would be more instructive to use the method in practice. 

\section{Related Work}
\label{sec:related-work}
\mypara{The Differences with the Existing Surveys} 
This study mainly differs from existing SoK papers~\cite{DBLP:journals/corr/abs-2109-10870, DBLP:journals/ijmir/RayR20, DBLP:conf/eurosp/PapernotMSW18, DBLP:journals/information/AsswadG21} in the following three aspects.  
1) Existing works usually discuss copyright issues from a model perspective rather than a dataset perspective. 
They focus on the theft and protection of model copyright, but ignore the copyright protection of training data. 
2) Image watermarking techniques for copyright protection, which are designed to defy traditional attacks~\cite{DBLP:conf/sswmc/KutterP99}, \eg, enhancement attacks, noise addition attacks, and compression attacks. 
However, traditional image watermarks can be readily removed by DNN models due to their remarkable feature extraction and generalization ability~\cite{DBLP:journals/spic/HatoumCCD21}. 
3) Existing works do not involve state-of-the-art techniques for copyright protection. 
For instance, the advanced dataset ownership resolution strategies, \eg, watermarking~\cite{DBLP:conf/eccv/ZouGW22} and data isotope~\cite{DBLP:journals/corr/abs-2208-13893}, are not included in~\cite{DBLP:journals/corr/abs-2109-10870}. 
Thus, we consider systematizing the novel image copyright auditing mechanisms optimized for DNN applications. 

\mypara{Differences and Relations between Dataset Copyright Auditing and Membership Inference Attacks}
Membership inference attacks (MIA)~\cite{CZWBHZ21, LWHSZBCFZ22} on DNN models aim to discern if a specific data sample is part of a model's training set, leveraging the model's predictive behavior, such as confidence levels. 
According to this property, the dataset owner can adopt membership inference to determine whether a specific sample of its dataset is used in the training of the suspicious model. 
Thus, \autoref{tab:overview-of-existing-work} includes several representative membership inference attacks suitable for the sample-level and user-level dataset copyright auditing. 

However, there exist some differences between MIA and dataset copyright auditing. 
The first distinction between dataset copyright auditing and membership inference attacks lies in their underlying assumptions. 
Dataset copyright auditing operates under the assumption that the auditor possesses comprehensive knowledge about the dataset being audited.
In contrast, membership inference attacks aim to minimize reliance on the target dataset.
Thus, dataset copyright auditing can effectively employ intrusive techniques, such as watermark injection, into the target dataset. 
Additionally, dataset copyright auditing methods typically analyze characteristics across a batch of samples for auditing purposes, whereas membership inference attacks are specifically tailored to assess individual samples. 

\section{Conclusion}
\label{sec:conclusion}
In this work, we evaluate the current state of dataset copyright auditing research and categorize existing methods into two categories: intrusive and non-intrusive, depending on the interaction with the original dataset. 
Then, we develop two frameworks to critically analyze the effectiveness of these methods in meeting the challenges posed by contemporary copyright issues. 
The analysis not only reviews existing methods, but also integrates findings from recent studies to provide a holistic view of the dataset copyright auditing landscape. 
We conclude with several promising directions for future research, which are necessary for auditing tools to fulfill the evolving requirements of effective copyright protection in machine learning applications. 
This work serves as a vital resource for practitioners, offering insights into the current state and potential advancements in the field of dataset copyright auditing. 
Beyond this aspect, we now outline avenues for future work to strengthen our understanding of dataset copyright auditing. 

\mypara{Limitation and Future Work}
The observations in this work focus mainly on the copyright audit of image datasets. 
However, there are many datasets in other domains that require copyright auditing, such as text, audio, tabular, and graph data, and corresponding dataset copyright auditing methods have been proposed in these fields. 
The frameworks proposed in this paper can still be applied for analysis in other domains. 
Thus, this paper can serve as a stepping stone for future research to systematize dataset copyright auditing tools across different domains. 

\section{Acknowledgements}
We would like to thank the shepherd and the anonymous reviewers for their constructive feedback. 
This work was partly supported by the National Natural Science Foundation of China (NSFC) under Grants No. 62293511, 62402379, U22A2029, U20A20175, and the Fundamental Research Funds for the Central Universities No. 226-2022-00107 and 226-2023-00111. 
Zhikun Zhang was supported in part by the NSFC under Grants No. 62402431 and Zhejiang University Education Foundation Qizhen Scholar Foundation. 
Min Chen was partly supported by the project CiCS of the research programme Gravitation which is (partly) financed by the Dutch Research Council (NWO) under the grant 024.006.037.

\balance
\bibliographystyle{IEEEtran}
\bibliography{refs}
\appendices

\section{Robustness Evaluation in Existing Studies}
\autoref{tab:ml-data-processing} shows the consideration of robustness evaluation in existing work, focusing on three processes: data preparation, model optimization, and model deployment. 
In addition to the operations depicted in \autoref{fig:ml_system}, prior studies also consider various adaptive adversarial settings, including output perturbation, neural cleanse, and anti-backdoor learning. 

\begin{table*}[h]
\caption{A summary of robustness evaluations considered in existing studies. 
Blank entries in the table indicate that the paper does not evaluate the impact of operations in this part on auditing performance. 
} 
\label{tab:ml-data-processing}
\centering
\setlength{\tabcolsep}{0.55em}
\renewcommand{\arraystretch}{1.1}
\begin{tabular}{c|c|c|c}
\toprule
\textbf{Method}                 & \textbf{Data Preparation}    &    \makebox[40ex][c]{\textbf{Model Optimization}}  &   
\makebox[40ex][c]{\textbf{Model Deployment}}
\\ \toprule
Sablayrolles~\etal~\cite{DBLP:conf/icml/SablayrollesDSJ20} &           &     &                           \\
Maini~\etal~\cite{DBLP:conf/iclr/MainiYP21} &    & Zero-short learning & Fine-tuning, Adversarial training            \\
Tang~\etal~\cite{DBLP:journals/corr/abs-2303-11470} & & & \\
Li~\etal~\cite{DBLP:journals/corr/abs-2209-06015} &  & Anti-backdoor learning & Fine-tuning, Model pruning \\
Zou~\etal~\cite{DBLP:conf/eccv/ZouGW22} & Data augmentation & & \\
\multirow{2}{*}{Wenger~\etal~\cite{DBLP:journals/corr/abs-2208-13893} } & & Transfer Learning &
Fine-tuning, Reducing outputs' granularity  \\
& & Adversarial augmentation &  Supurious correlation detection, Feature inspection\\
Choquette{-}Choo~\etal~\cite{DBLP:conf/icml/Choquette-ChooT21} & & Differential privacy, L2-norm regularization & Simplifying model's confidences  \\
Li~\etal~\cite{LZ21}  & & & MemGuard, Adversarial regularization \\
Xu~\etal~\cite{Xu2022DataOI} & & & \\
Chen~\etal~\cite{CZWBZ23} &Input perturbation & DP-SGD & Output perturbation \\
Dong~\etal~\cite{DBLP:conf/ndss/DongLCXZ023} & Data augmentation & Adversarial fine-tuning & Static modification \\
Li~\etal~\cite{DBLP:conf/nips/LiBJYXL22} &  &  & Fine-tuning, Model pruning \\
\multirow{2}{*}{Shokri \etal ~\cite{SSSS17}} & \multirow{2}{*}{ }& \multirow{2}{*}{L2-norm regularization} & Top-K selection, Prediction quantization\\
& & & Entropy enhancement\\
Sablayrolles~\etal~\cite{SDSOJ19} & & &  \\
Salem~\etal~\cite{SZHBFB19} & & Dropout, Ensemble learning & \\
Leino~\etal~\cite{LF20} & & Differential privacy, Dropout & \\
Sommer~\etal~\cite{SSWM20} & & & Neural cleanse \\
Li~\etal~\cite{DBLP:journals/corr/abs-2010-05821} & & & \\
Song~\etal~\cite{SM21}  & & Adversarial regularization, Early stopping & \\
Hu~\etal~\cite{DBLP:conf/ijcai/HuSDCSZ22} & & Differential privacy & Neural cleanse \\
Liu~\etal ~\cite{liu2022membership} & & & \\
Liu~\etal~\cite{liu2023watermarking} & & & \\
Guo~\etal~\cite{guo2023domain} & Data augmentation &Domain adaption & Fine-tuning, Model pruning, Neural cleanse\\
Tekgul~\etal~\cite{tekgul2022effectiveness} & Data augmentation & &  \\
Dziedzic~\etal~\cite{DBLP:journals/corr/abs-2209-09024} & & Shuffle, Padding drops& \\
Li~\etal~\cite{li2023functionmarker} & & & Watermark detection attack, Watermark rewrite attack \\
Li~\etal~\cite{li2023black} & & & Fine-tuning, Model pruning, Anti-backdoor learning \\
\bottomrule 
\end{tabular}
\end{table*}

\clearpage

\section{Meta-Review}
The following meta-review was prepared by the program committee for the 2025 IEEE Symposium on Security and Privacy (S\&P) as part of the review process as
detailed in the call for papers. 

\subsection{Summary}
The paper provides a Systematization of Knowledge (SoK) on dataset copyright auditing in machine learning systems. 
It categorizes existing methods into intrusive and non-intrusive approaches, assessing their effectiveness and limitations. 
The authors analyze various methods within the context of real-world applications, provide empirical evaluations on selected approaches, and identify gaps in the research. 
Future directions for developing robust auditing tools are also proposed, with a focus on improving practical deployment and copyright protection in machine learning pipelines. 

\subsection{Scientific Contributions}
\begin{itemize}
    \item Independent confirmation of important results with limited prior research
    \item Provides a valuable step forward in an established field
\end{itemize}

\subsection{Reasons for Acceptance}
\begin{enumerate}
    \item The paper addresses a highly relevant issue in machine learning regarding unauthorized data use, making it valuable for the community. 
    \item The paper offers important takeaways that will benefit both researchers and practitioners in the field of machine learning and copyright protection.
\end{enumerate}

\end{document}